\documentclass[twocolumn,amssymb,prb]{revtex4}
\usepackage{epsfig}
\usepackage{graphicx}

\begin{document}
\title{Sub-gap Fano resonances in a topological superconducting wire with on-site Coulomb interactions}

\author{Piotr Stefa\'nski}
\email{piotrs@ifmpan.poznan.pl}
 \affiliation{Institute of Molecular Physics of the Polish Academy of Sciences\\
  ul. Smoluchowskiego 17, 60-179 Pozna\'n, Poland}
\author{}
\affiliation{}
\author{}
\affiliation{}
\date{\today}

\begin{abstract}
We consider theoretically a $1D$-semiconducting wire with strong Rashba interaction in proximity with $s$-wave superconductor, driven into topological phase by external magnetic field. Additionally, we take into account on-site Coulomb interactions inside the wire. The system is modelled by a tight binding Hamiltonian with  Rashba hopping term and induced $s$-wave superconductivity. Calculations are performed utilizing recursive Green's function method, and Coulomb interactions are treated selfconsistently within Hubbard $I$ approximation. For the Hubbard levels residing within $p$-wave superconducting gap, particle-hole symmetric four-resonance structure develops in the density of states, apart from Majorana resonance. One pair of particle-hole symmetric resonances is created by the discrete $II$-Hubbard levels of the particular site, and the second pair of Hubbard sub-bands originates from recursive summation over the sites of the wire. Quantum interference between both types of pairs of states creates in-gap charge-conjugated Fano resonances with opposite asymmetry factors. We demonstrate that when quantum interference is dominated by two-particle tunneling, the Majorana resonance is strongly diminished, while it is not altered when single-particle tunneling dominates in interference process. We also discuss some consequences for experimental distinction of true Majorana states, and show that on-site Coulomb interactions support the appearance of topological phase.
\end{abstract}
\maketitle

\section{Introduction}
Majorana fermions, exotic quantum particles being their own antiparticles, were proposed by Ettore Majorana \cite{EttoreMajorana1937} as a real solution of Dirac equation. Recently their realization in solid state has been theoretically predicted, and then it was realized experimentally \cite{DasSarma2015,Aguado2017a}. In solid state heterostructures they appear as quasiparticles, so called Majorana zero-modes (MZM)\cite{Beenakker2019}. The quest for creating them in solid state is not only due to their exotic properties per se, but also due to the possibility of performing  with them logical operations, free of local decoherence processes \cite{Nayak2008,Stanescu2017book}. The most promising for Majorana braiding operations are MZM at the ends of a topological wire \cite{Kitaev2000b,Lutchyn2010b,Oreg2010a}, which is realized in heterostructures comprising a quasi-$1D$ semiconducting wire with strong spin-orbit (SO) interaction proximitized to $s$-wave superconductor. This setup was implemented experimentally by Mourik \textit{ et.al.} \cite{Mourik2012} and further realizations were continued by other researchers \cite{Das2012,Deng2014a,Higginbotham2015,Chang2015,Deng2016a,Chen2017a,Moor2018,Bommer2019}. When subjected to external magnetic field perpendicular to SO Rashba field, the wire enters topological state with effective $p$-wave pairing.

Majorana zero-modes emerging in solid state are inevitably subjected to the influence of various processes present in this  environment. Beside of decoherence processes \cite{Karzig2021}, the influence of Coulomb interactions and quantum interference on MZM formation is of the most importance.

Coulomb interactions between electrons are fundamental and unavoidable phenomenon in solid state. They modify various properties of materials in a dramatic way. The most spectacular examples are Hubbard-Mott transition \cite{Gebhard1997} and Kondo effect \cite{Hewson1993}, which gained a new insight and control of parameters when reproduced in nanodevices \cite{Kastner1998}.

Majorana quasiparticles and generally topological superconducting state in presence of Coulomb interactions  gained large interest and were investigated  theoretically. In Kitaev chains, the paradigm of topological superconductor, the influence of nearest neighbor interactions were investigated \cite{Gangadharaiah2011,Katsura2015,Ng2015b,Herviou2016a,Miao2017a,Sekania2017,Ezawa2017,Li2018c,Miao2018,Sarkar2020,Zvyagin2021}, focusing on the change of the ground state and parity of the model, decoherence and braiding.
The influence of the on-site Coulomb interactions on the topological phase diagram was also intensively investigated by various methods and models, proving that these interactions promote the emergence of topological state \cite{Stoudenmire2011b,Lutchyn2011,Klinovaja2012,Maier2014,Meidan2014,Manolescu2014,Kells2015b,Chan2015,Zhang2015a,Schmidt2016,
Xu2016,Thakurathi2020,Rylands2020,Wang2020,Mahyaeh2020,Aksenov2020}.
Charging effects in electron transport were also investigated in so-called Coulomb islands made of the section of topological wire \cite{Hutzen2012a,Vijay2016,Lu2016,Lutchyn2017,Chiu2017,Hell2018,Glazman2019}.

Interestingly, it was also demonstrated that Coulomb interactions can induce topological transition in the system with no spin-orbit interaction\cite{Haim2016,Li2019,Zhu2021}, as they mix the spins of interacting electrons.

Quantum interference as a characteristic manifestation of quantum mechanics, has also gained renewed interest when observed in controlled environment in nanostructures \cite{Miroshnichenko2010}. Interestingly, interpretation of the spectra of ionized atoms was also in the field of interest of Majorana \cite{Majorana1931}, and then the theory of characteristic antisymmetric resonances in the atomic spectra has been further developed by Fano \cite{Fano1961}, whose work is a direct continuation of Majorana findings \cite{Vittorini-Orgeas2009}.

Fano resonances  in the presence of MZMs were investigated in various transmission geometries, allowing quantum interference between multiple propagation paths, one of them being the Majorana channel\cite{Gong2014,Dessotti2014,Ueda2014,Gong2016b,Nesterov2015,Baranski2017,Schuray2017a,Ramos-Andrade2019,Calle2020,Gong2021}. The purpose of such approaches was mainly identification of Majorana states due to their chirality, manifested when MZM is tunnel-coupled to external reservoirs.

In the present paper we demonstrate, how the coexistence of quantum interference and Coulomb interactions in a $1D$-topological wire influences the formation of Majorana resonance in the density of states. When the discrete Hubbard levels at each site enter the superconducting gap, they acquire their charge-conjugated hole counterparts. The appearance of the particle-hole symmetric Hubbard resonances at each site  produces \textit{globally} quasiparticle Hubbard sub-bands in the density of states of the wire.  The \textit{local} Hubbard quasiparticle resonances of a particular site are also visible in the density of states of the wire, when calculated at a particular site. The end-sites of the wire are of special importance, because the MZM wave functions localize there, when the wire is driven into topological state. We demonstrate that, as a result of quantum interference between the particle-hole symmetric pairs of local in-gap quasiparticle Hubbard resonances and pairs of in-gap quasiparticle Hubbard sub-bands, characteristic particle-hole symmetric Fano resonances with opposite $q$-asymmetry parameters, form inside the topological superconducting gap. This quantum interference is mediated by propagation of superconditing pairs at Fermi energy, and it influences the formation of the Majorana resonance. When two-particle tunneling processes between local Hubbard levels and Hubbard bands dominate, the Majorana resonance is diminished considerably, whereas for the dominance of one-particle tunneling processes the Majorana resonance remains unaltered. The dominance of the particular type of tunneling, in turn, depends on the relative positions of the interfering sub-gap levels; two-particle tunneling dominates when the interfering levels are positioned in the opposite charge-conjugated sectors: particle or hole, and single-particle tunneling dominates when both the interfering levels are in the same sector. Eventually, the Majorana resonance vanishes completely when a pair of particle-hole local Hubbard levels is shifted into resonance with Fermi energy by a magnetic field.

The paper is organized as follows. The Hamiltonian of a wire with large spin-orbit coupling and in proximity of $s$-wave superconductor, as well as the description of the calculation method, is presented in Section II. In Section III the numerical results are presented, and two toy models are introduced for their interpretation, followed by discussion and suggestions for an experimental realization. The last section includes conclusions, and the details of calculations are presented in Appendices.

\section{The model and calculation approach}
The semiconducting wire with large spin-orbit interaction is modelled by the tight-binding Hamiltonian with on-site local Coulomb interactions. The wire is in proximity to $s$-wave superconductor, which induces  superconducting pairing in it. Additionally, the wire is subjected to an external magnetic field, which can drive the wire into topological state. Coulomb interactions at each site of the wire are treated within Hubbard $I$ approximation; as a result two  Hubbard resonances (per spin), separated by the Coulomb repulsion $U$, form in the density of states of each site.

The wire is set along $x$-direction, subjected to external magnetic field, $V_{z}$, in $z$-direction, perpendicular to spin-orbit Rashba field. It is described by the Hamiltonian \cite{Stoudenmire2011b,Huang2014a} : $H_{wire}=H_{0}+H_{so}+H_{sc}+H_{int}$, where:

\begin{eqnarray}\label{Htight}
H_{0}=\sum_{j=1}^{N}\sum_{\sigma=\downarrow,\uparrow}\epsilon_{\sigma}c_{j\sigma}^{\dagger}c_{j\sigma}
-t\sum_{j=1,\sigma}^{N-1}\left(c_{j+1\sigma}^{\dagger}c_{j\sigma}+h.c. \right)\\
H_{so}=\sum_{j=1}^{N-1}\sum_{\sigma,\sigma'}(-i t_{so})c_{j+1\sigma}^{\dagger}\hat{\sigma}^{y}_{\sigma\sigma'}c_{j\sigma'}+h.c.\\
H_{sc}=\Delta\sum_{j=1}^{N}\left(c_{j\uparrow}^{\dagger}c_{j\downarrow}^{\dagger}+c_{j\downarrow}c_{j\uparrow} \right)\\
H_{int}=\sum_{j=1}^{N}U n_{j\downarrow}n_{j\uparrow},
\end{eqnarray}
where $\epsilon_{\downarrow/\uparrow}=-(\mu-2t)\mp V_{z}$.
$H_{0}$ describes tight-binding part of the Hamiltonian with $t=\hbar^{2}/(2 m^{\star} a^{2})$-nearest neighbor hopping amplitude between the sites, with chemical potential $\mu$, subjected to the magnetic field, $m^{\star}$ being the effective electron mass and $a$ the lattice constant. The operator $c_{j\sigma}^{\dagger}$ ($c_{j\sigma}$) creates (annihilates) an electron of the spin $\sigma$ at the site $j$ of the wire. $H_{so}$ describes the effect of spin-orbit Rashba coupling with $t_{so}=\sqrt{E_{so}t}$, $E_{so}=m^{\star}\alpha^{2}/(2\hbar^{2})$, and $\alpha$- the spin-orbit coupling strength in the wire \cite{Rainis2013}. $H_{sc}$ describes induced superconducting pairing with amplitude $\Delta$, assumed to be real. Finally, $H_{int}$ describes on-site Coulomb interactions.

In our numerical studies we assumed the tight-binding hopping to be the energy unit, and relations between other parameters have been chosen to favor topological phase \cite{Stoudenmire2011b,Rainis2013}: $t\gg t_{so}>\Delta$ with $\Delta=0.2$, and $t_{so}=2\Delta$, and chemical potential $\mu=1$. The wire has been assumed to have the length of $N=300$ sites. For the hopping amplitude $t=10 meV$, the lattice constant $a=15 nm$, which yields the wire length $L=4.5 \mu m$, comparable to the wire dimension investigated experimentally \cite{Mourik2012}.  The topological phase is induced by the increase of the magnetic field, and exists for fields above the critical value $|V_{z}|>V_{z}^{cr}=\sqrt{\mu^{2}+\Delta^{2}}$  \cite{Lutchyn2010b,Oreg2010a}. The on-site Coulomb interactions inside the wire are treated within Hubbard $I$ approximation. Their main effect is the appearance of additional sub-gap states, which, as we demonstrate, can have substantial effect on the MZM formation.

Details of calculations can be found in Appendix \ref{AppendixA}.

\section{Results and discussion}
\subsection{Majorana resonance in presence of in-gap Fano resonances}
To study our system we calculate the density of states of the wire in topological state at it's end-site $i=1$, $\rho(\omega)=\sum_{\sigma=\downarrow,\uparrow}\rho_{\sigma}(\omega)=-(1/\pi)Im[[\hat{G}_{1,1}(\omega)]_{1,1}+[\hat{G}_{1,1}(\omega)]_{2,2}]$ (see Appendix \ref{AppendixA}). Density of states can easily be measured experimentally by tunneling spectroscopy; examining differential conductance between spin-polarized STM tip and the wire vs. energy and tip distance \cite{Jeon2017a}, or between the topological wire and a normal metal electrode \cite{Prada2020}.

The Hubbard $I$ approximation for on-site Coulomb interactions has a static effect of the appearance of Hubbard resonances $\epsilon^{I}_{\sigma}$ and $\epsilon^{II}_{\sigma}$, separated by Coulomb repulsion $U$. For the discussion of the results, it is worth to analyze the sequence of Hubbard levels which arise at each site $i$. Caused by induced superconductivity present  in the wire, the particle $(p)$ Hubbard levels acquire their charge-conjugated hole $(h)$ counterparts.  Their $(p)$-location follows from the poles of the diagonal matrix elements $(1,1)$ and $(2,2)$ of $\hat{g}_{0}$, Eq.~(\ref{g0int}):
\begin{eqnarray}
\epsilon^{Ip}_{i\downarrow}=\epsilon_{i\downarrow}=-(\mu-2t)-V_{z},\\
\epsilon^{IIp}_{i\downarrow}=\epsilon_{i\downarrow}+U=-(\mu-2t)+U-V_{z},\\
\epsilon^{Ip}_{i\uparrow}=\epsilon_{i\uparrow}=-(\mu-2t)+V_{z},\\
\epsilon^{IIp}_{i\uparrow}=\epsilon_{i\uparrow}+U=-(\mu-2t)+U+V_{z}.
\end{eqnarray}
Location of the $h$-levels follows form the poles of the diagonal matrix elements $(3,3)$ and $(4,4)$ of $\hat{g}_{0}$ and is particle-hole symmetric with respect to $p$-levels: $\epsilon^{Kh}_{i\sigma}=-\epsilon^{Kp}_{i\sigma}$, $K=I,II$ and $\sigma=\downarrow,\uparrow$.

For the parameters utilized in the numerical calculations, the profound influence on the Majorana resonance oroginates from the $II$-nd Hubbard level $\epsilon^{IIp}_{i\downarrow}$ and its counterpart $\epsilon^{IIh}_{i\downarrow}$; shifted by the magnetic field into the topological superconducting gap. Note that in the topological phase the index $\downarrow$ should be regarded as chiral index of the active sub-band, distinguished by the direction of the external magnetic field \cite{Alicea2010,Sau2010a,Alicea2012b}, in our case $V_{z}>0$.

We present the results of the density of states of the wire for the on-site Coulomb repulsion $U=2.5\Delta$ and magnetic field $V_{z}>V_{z}^{cr}$; the wire being in topological phase. $V_{z}^{\star}$ denotes the value of magnetic field by  for which $\epsilon^{IIp}_{\downarrow}=-\epsilon^{IIh}_{\downarrow}=\epsilon_{F}$. As we demonstrate below, for $V_{z}<V_{z}^{\star}$ and for $V_{z}>V_{z}^{\star}$ very distinct impact on MZM resonance can be observed when the sub-gap levels are shifted by magnetic field.

\begin{figure}
  \centering
  \includegraphics[width=9cm]{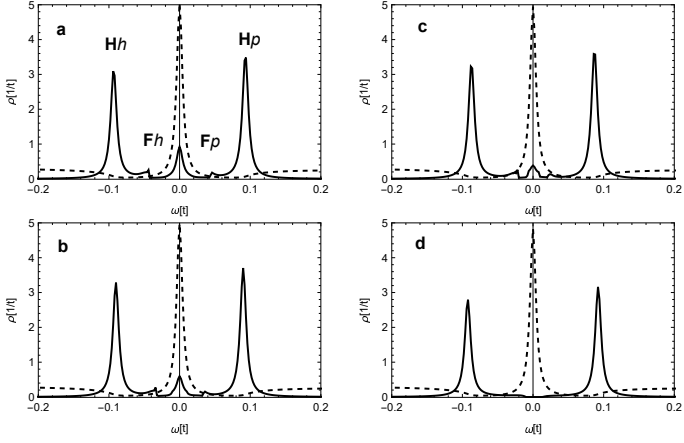}
  \caption{Density of states calculated at the site $i=1$ for magnetic field $V_{z}\leq V_{z}^{\star}$, increasing from $(a)$ to $(d)$ - solid lines. Dashed curves represent the curves calculated for the same parameters but $U=0$. Panel $a$- $V_{z}=1.42 V_{z}^{cr}$, Panel $b$ - $V_{z}=1.43 V_{z}^{cr}$, Panel $c$ - $V_{z}=1.44 V_{z}^{cr}$ and Panel $d$ - $V_{z}=V_{z}^{\star}=1.47 V_{z}^{cr}$. The curves are calculated for $t=1$, $\mu=1$, $\Delta=0.2$, $t_{so}=0.4$ and $U=2.5\Delta$.} \label{specdensbelow}
\end{figure}

Fig.~(\ref{specdensbelow}) displays density of states calculated  recursively for the first site of the wire in the topological phase, for  magnetic field $V_{z}<V_{z}^{\star}$ increasing from $(a)$ to $(d)$. As a consequence of entering of the charge-conjugated pair of Hubbard levels $\epsilon^{IIp}_{i\downarrow}$ and $\epsilon^{IIh}_{i\downarrow}$ into the superconducting gap, the density of states has a richer structure as compared to noninteracting case (dashed lines). Apart from Majorana resonance pinned at Fermi energy, two pairs of particle-hole symmetric peaks are observed in the density of states. The large and broad pair is a result of recursive summation over the sites of the wire from $i=2$ to $N$, each with $\epsilon^{IIp}_{i\downarrow}$ and $\epsilon^{IIh}_{i\downarrow}$ pair, and form two Hubbard sub-bands  marked as H\textit{h} and H\textit{p}. Another pair of resonances, with characteristic asymmetric shape, arise from the local pair of  $\epsilon^{IIp}_{1\downarrow}$ and $\epsilon^{IIh}_{1\downarrow}$ of the first site,  marked as F\textit{h} and F\textit{p} in the picture. The asymmetric Fano shape is caused by quantum interference of these discrete sites with Hubbard sub-bands, and the relation of the Fano asymmetry parameters of charge-conjugated resonances is $q_{p}=-q_{h}$. The mechanism of appearance of these resonances is discussed in detail within Toy Model $I$.

As the magnetic field increases, and shifts the pair $\epsilon^{IIp/h}_{1\downarrow}$ towards Fermi energy, the Majorana peak is strongly diminished up to the magnetic field value $V_{z}=V_{z}^{\star}$, when it is destroyed complectly for $\epsilon^{IIp/h}_{1\downarrow}=\epsilon_{F}$. The characteristic diminishing of the Majorana peak, as demonstrated in Toy Model $I$, is caused by two-particle dominated tunneling on Fermi energy between discrete $\epsilon^{IIp/h}_{1\downarrow}$ levels and Hubbard sub-bands. For $V_{z}<V_{z}^{\star}$ the tunneling takes place between discrete $i=1$-site Hubbard level $\epsilon^{IIp}_{1\downarrow}$ ($\epsilon^{IIh}_{1\downarrow}$) positioned in the particle (hole) sector and the Hubbard sub-band positioned in hole (particle) sector.

\begin{figure}
  \centering
  \includegraphics[width=9cm]{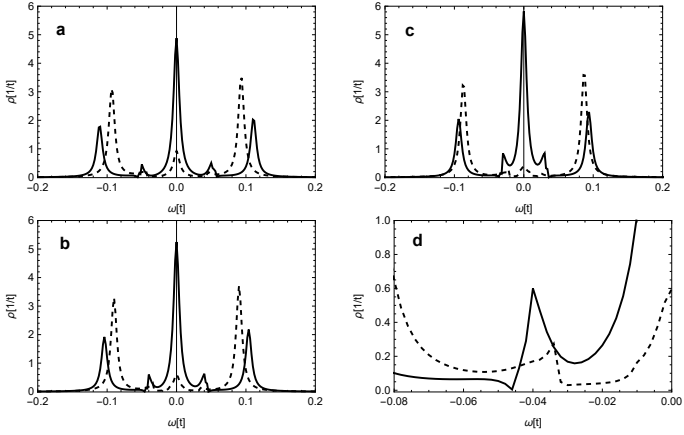}
  \caption{Density of states calculated at the site $i=1$ for magnetic field $V_{z} > V_{z}^{\star}$, decreasing from $(a)$ to $(c)$- solid lines. Dashed curves represent respective spectral densities from  panels $(a)$ to $(c)$ in Fig.~(\ref{specdensbelow}), which correspond to the magnetic fields when the exchange in positions between $\epsilon_{\downarrow}^{IIp}$ and  $\epsilon_{\downarrow}^{IIh}$ takes place. Panel $a$ - $V_{z}=1.52 V_{z}^{cr}$, Panel $b$ - $V_{z}=1.51 V_{z}^{cr}$, Panel $c$ - $V_{z}=1.50 V_{z}^{cr}$. In Panel $d$ magnified Fano resonances from Panel $b$ are depicted.  The curves are calculated for $t=1$, $\mu=1$, $\Delta=0.2$, $t_{so}=0.4$ and $U=2.5\Delta$.}\label{specdensabove}
\end{figure}

An interesting feature emerges when the magnetic field is increased further, above $V_{z}^{\star}$ value, which is demonstrated in Fig.~(\ref{specdensabove}). For $V_{z}>V_{z}^{\star}$ the second Hubbard levels $\epsilon^{IIp}_{1\downarrow}$ and $\epsilon^{IIh}_{1\downarrow}$ exchange their positions in energy scale, as compared to the corresponding fields for $V_{z}<V_{z}^{\star}$.  In panels $(a)$ to $(c)$, the dashed curves are the same as in Fig.~(\ref{specdensbelow}) $(a)$-$(c)$ for the corresponding fields $V_{z}<V_{z}^{\star}$. Strikingly, the Majorana resonance is not diminished for $V_{z}>V_{z}^{\star}$.
Although with the shift by the magnetic field, the discrete Hubbard levels cross Fermi energy and exchange their positions, the Hubbard sub-bands do not change their positions. This feature is demonstrated in Toy Model $I$, where the exchange of the position of the impurity $\epsilon_{i}\rightarrow -\epsilon_{i}$ does not alter the density od states of the superconductor.  Thus, in this regime quantum interference takes place between discrete Hubbard levels, which have exchanged their positions and the broad Hubbard sub-bands which did not change their positions.  This results in the quantum interference between discrete Hubbard levels and the Hubbard sub-bands positioned in \textit{the same} particle or hole sector. As shown in Toy Model $I$, for such configuration  single-particle tunneling dominates, which has negligible effect of MZM resonance.

For $V_{z}=V_{z}^{\star}$ the Hubbard levels $\epsilon^{IIp}_{1\downarrow}=\epsilon^{II}_{1\downarrow}=\epsilon_{F}$, which results in a complete destruction of the Majorana resonance. This process is analyzed in Toy Model $II$; in the case of direct tunneling into Majorana zero mode, the processes of single-particle and two-particle tunneling have the same contributions.

\begin{figure}
  \centering
  \includegraphics[width=7cm]{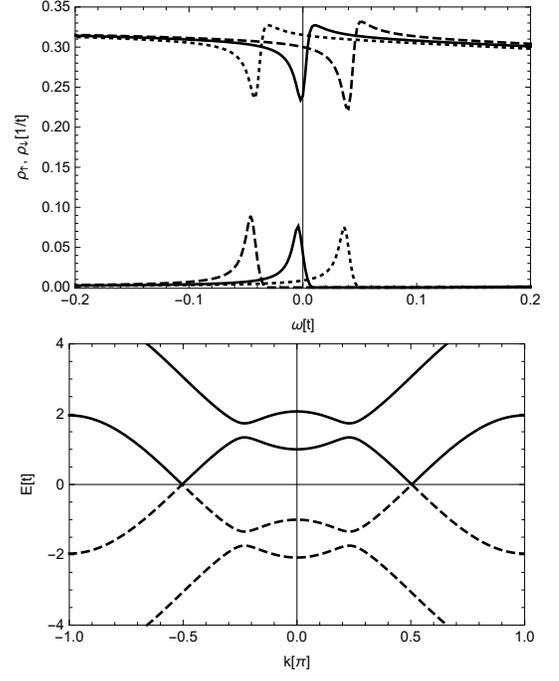}
  \caption{Upper Panel: spectral densities of the wire at the site $i=1$, calculated for the same parameters as in Figs.~(\ref{specdensbelow}) and (\ref{specdensabove}), but for $t_{so}=0$ trivial phase. The upper curves represent $\rho_{\downarrow}$ and the lower curves represent $\rho_{\uparrow}$. The dashed curves are for  $V_{z}=1.43 V_{z}^{cr}$, solid curves are for  $V_{z}= V_{z}^{\star}$  and dotted for  $V_{z}=1.51 V_{z}^{cr}$. The lower Panel displays the energy spectrum of the wire for periodic model in momentum space for the same parameters and $V_{z}=V_{z}^{\star}$.}\label{specdenstso0}
\end{figure}

\textit{The wire in its trivial state}.
Let us analyze for comparison the density of states of the wire at site $i=1$ in its trivial state,  by assuming the absence of spin-orbit interaction, $t_{so}=0$. The results are presented in the upper Panel of Fig.~(\ref{specdenstso0}). The upper (lower) curves in this Panel represent spin-down (spin-up) spectral densities calculated from $\rho_{\downarrow}(\omega)=-(1/\pi)Im[\hat{G}_{1,1}(\omega)]_{1,1}$ ($\rho_{\uparrow}(\omega)=-(1/\pi)Im[\hat{G}_{1,1}(\omega)]_{2,2}$).  The lower Panel of  Fig.~(\ref{specdenstso0}) depicts the energy spectrum of the wire for the periodic model, calculated from Eqs~(\ref{E1/2}) and (\ref{E3/4}). In the absence of spin-orbit interaction the spin quantum number becomes a conserved quantity. The superconducting gap is closed at finite momentum by touching of the lower particle and the higher hole bands (solid and dashed curves, respectively) at Fermi energy. The Fano resonance, with $q_{p}>0$, visible in spin-down density of states, arises as a result of quantum interference between the discrete second Hubbard level $\epsilon_{1\downarrow}^{IIp}$ of the first site with the lower quasiparticle band (solid curve in the lower Panel). This band originates from spin-down band of the wire, in the absence of $s$-wave correlations. The resonance is shifted by the external magnetic field. At the same time the  Fano resonance, with $q_{h}< 0$, arises  due to quantum interference of the hole second Hubbard resonance $\epsilon_{1\downarrow}^{IIh}$ with the higher hole band (dashed curve in the lower Panel), and is shifted by the magnetic field in the opposite direction with respect to the particle Fano resonance. Its evolution is visible in the spin-up density of states because in the presence of the s-wave ordering the hole quasiparticle band exhibits the majority of up spins as opposed to its particle counterpart with spin-down majority. When the wire is driven into topological phase by switching on large spin-orbit interaction, both particle and hole Fano resonances appear in the density of states of the common chirality, determined by the direction of external magnetic field.

\subsection{Modification of the critical field by Coulomb interactions}
To examine the influence of Coulomb interactions on the topological phase transition we perform the transformation into $k$-space of the original Hamiltonian and examine its spectrum. In the first step we perform the transformation of the noninteracting Hamiltonian $H_{0}+H_{so}+H_{sc}$ of Eq.~(\ref{Htight}).

After assuming closed periodic boundary conditions we take the expressions of the transformed operators for site $j$: $c_{j,\sigma}=(1/\sqrt{N})\sum_{k}\exp(-i k x_{j})c_{k,\sigma}$ and the representation of the Dirac delta function: $\delta_{k,k'}=(1/N)\sum_{j=1}^{N}\exp[i(k-k')x_{j}]$. The transformed Hamiltonian assumes the form:
\begin{eqnarray}
\nonumber
H_{0}=\sum_{k,\sigma}[\epsilon_{\sigma}-2t\cos(k a)]c^{\dagger}_{k\sigma}c_{k\sigma},\\
H_{R}=2 i t_{so}\sum_{k}(c^{\dagger}_{k\uparrow}c_{k\downarrow}-c^{\dagger}_{k\downarrow}c_{k\uparrow})\sin(k a),\\
\nonumber
H_{sc}=\sum_{k}(\Delta c_{k\downarrow}c_{-k\uparrow}+\Delta^{\star}c_{-k\uparrow}^{\dagger}c_{k\downarrow}^{\dagger}),
\end{eqnarray}
where $\epsilon_{\sigma}=-(\mu-2 t)\mp V_{z}$ for spin $\sigma=\downarrow$ and $\sigma=\uparrow$, respectively.

In the next step we rewrite the Hamiltonian in the Nambu basis by introducing spinor $\Psi=(c_{k\downarrow}, c_{k\uparrow},c^{\dagger}_{-k\uparrow},c^{\dagger}_{-k\downarrow} )$, and diagonalize BdG Hamiltonian matrix:
\begin{equation}\label{H_BdG}
  H=\frac{1}{2}\sum_{k}\Psi^{\dagger}H_{BdG}\Psi+\frac{1}{2}\sum_{k,\sigma}[\epsilon_{\sigma}-2t\cos(ka)],
\end{equation}
\begin{eqnarray}
H_{BdG}=
\left(
\begin{array}{cccc}
\tilde{\epsilon}_{\downarrow} & -\tilde{t}_{so} & -\Delta^{\star} & 0\\
\tilde{t}_{so} & \tilde{\epsilon}_{\uparrow} & 0 & \Delta^{\star}\\
-\Delta & 0 & -\tilde{\epsilon}_{\uparrow} & -\tilde{t}_{so}\\
0 & \Delta & \tilde{t}_{so} & -\tilde{\epsilon}_{\downarrow}
\end{array}
\right).
\end{eqnarray}

Here, we have introduced  the abbreviations $\tilde{\epsilon}_{\sigma}=\epsilon_{\sigma}-2t\cos(ka)$ and $\tilde{t}_{so}=2i t_{so}\sin(ka)$.
Diagonalization of $H_{BdG}$ matrix uncovers the following sub-bands:
\begin{eqnarray}
\label{E1/2}
  E_{1/2}=\mp\frac{1}{\sqrt{2}}(\sqrt{A-\sqrt{B}}) \\\label{E3/4}
  E_{3/4}=\mp\frac{1}{\sqrt{2}}(\sqrt{A+\sqrt{B}})\\\nonumber
  A=\tilde{t}_{so}^{2}+ \tilde{\epsilon}_{\uparrow}^{2}+\tilde{\epsilon}_{\downarrow}^{2}+2\Delta^{2}\\\nonumber
  B=4\tilde{t}_{so}^{2}(\tilde{\epsilon}_{\uparrow}+\tilde{\epsilon}_{\downarrow})^{2}+
  (\tilde{\epsilon}_{\uparrow}-\tilde{\epsilon}_{\downarrow})^{2}
  [(\tilde{\epsilon}_{\uparrow}+\tilde{\epsilon}_{\downarrow})^{2}+4\Delta^{2}].
\end{eqnarray}
\begin{figure}
  \centering
  \includegraphics[width=8cm]{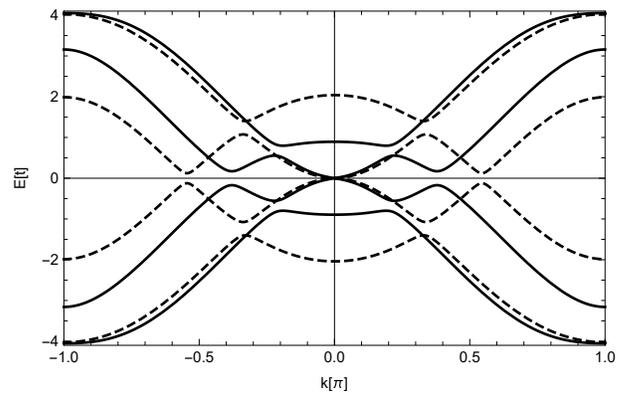}
  \caption{Energy spectrum calculated for noninteracting case for $V_{z}=V_{z}^{cr0}$ -dashed lines; the lower particle sub-band and the highest hole sub-band close the gap for $k=0$. Solid lines are calculated for $U=3\Delta$ and $V_{z}=0.44 V_{z}^{cr0}$; the gap is closing lower fields as compared to $U=0$ case. The dependencies are calculated for $t=1$, $\mu=1$, $t_{so}=0.4$, $\Delta=0.2$ and the lattice constant taken $a=1$.}\label{Gap_close}
\end{figure}

The critical magnetic field, at which the $s$-wave superconducting gap closes, follows from the relation $A=\sqrt{B}$ when the lowest (highest) sub-band from particle (hole) sector, Eq.~(\ref{E1/2}), touches Fermi level, as depicted  in Fig.~(\ref{Gap_close}) by dashed lines, and has the value $V_{z}^{cr0}=\sqrt{\mu^{2}+\Delta^{2}}$. Then the gap reopens as $V_{z}$ increases and the system enters topological superconducting phase.

In order to demonstrate, how the presence of on-site electron correlations modify the opening the $p$-wave superconducting gap we apply Hartree-Fock (HF) approximation to the interacting term in Hamiltonian Eq.~(\ref{Htight}): $Un_{j\downarrow}n_{j\uparrow}\rightarrow U\langle n_{j\uparrow}\rangle n_{j\downarrow}+U\langle n_{j\downarrow}\rangle n_{j\uparrow}+const$. As a result, each localized level is renormalized by Coulomb interaction: $\epsilon_{j\sigma}\rightarrow\epsilon_{j\sigma}+\langle n_{j\bar{\sigma}}\rangle U$. It is worth noticing that for the empty level, $\langle n_{j\sigma}\rangle=0$, and for the fully occupied level,  $\langle n_{j\sigma}\rangle=1$,  HF approximation is equivalent to Hubbard $I$ approximation, thus the discussion of these two limiting cases is consistent with our approximation used for numerical calculations. Namely, for $\langle n_{j\sigma}\rangle=0$ in both HF and Hubbard $I$ approximations the total spectral weight in the density of states is shifted to the bare $\epsilon_{j\sigma}$ level, whereas for high charge density, $\langle n_{j\sigma}\rangle=1$, and dominance of electron interactions the total spectral weight is shifted to $\epsilon_{j\sigma}+U$ level. The first case in equivalent to noninteracting model, and the second case, fully interacting, is easily obtained from the noninteracting model by the renormalization $\epsilon_{j\sigma}\rightarrow\epsilon_{j\sigma}+U$. When substituting $\tilde{\epsilon}_{\sigma}\rightarrow\tilde{\epsilon}_{\sigma}+U$ in Eqs.~(\ref{E1/2}) and (\ref{E3/4}), the $s$-wave gap closes for much lower magnetic field in comparison to the noninteracting case, as demonstrated in Fig.~(\ref{Gap_close}) by solid lines. Examining the condition for the gap closing we obtain the equation for the critical field modified by Coulomb interactions: $V_{z}^{crU}=\sqrt{(\mu-U)^{2}+\Delta^{2}}$. Thus, on-site Coulomb interactions promote the appearance of topological phase, a feature beneficial from the experimental standpoint. Similar conclusions were drawn from HF analysis and density matrix renormalization group approach\cite{Stoudenmire2011b}.

\subsection{Toy Models}
To achieve better understanding of our numerical results we introduce two Toy Models. In both the models we partition the wire into the first site $i=1$ and  the rest of the wire from $i=2$ to $i=N$, being in topological phase, mutually coupled. Within Toy Model $I$ we interpret the appearance of in-gap Fano resonances as a result of quantum interference between discrete Hubbard levels of the first site of the wire and Hubbard sub-bands. We emphasize the substantial difference in the impact of these interference processes on MZM resonance when two-particle tunneling dominates, as compared to the case of dominating single particle tunneling. In  Toy Model $II$ we analyze the destruction of MZM resonance by direct tunneling to the discrete Hubbard level of the first site.
\subsubsection{Toy Model $I$: Fano resonances as in-gap states  of topological superconductor}
\textit{Emergence of Hubbard sub-bands}.
Within this model the rest of the wire is described by the $p$-wave Hamiltonian with embedded impurity. The simplest version of the $p$-wave Hamiltonian  in momentum space reads \cite{Alicea2012b}:
\begin{eqnarray}\label{Ham_kitaev}\nonumber
H_{p}=\sum_{k}\epsilon_{k\alpha}c_{k\alpha}^{\dagger}c_{k\alpha}+\\
\sum_{k}[\frac{1}{2}\Delta_{p}(k)c_{k\alpha}^{\dagger}c_{-k\alpha}^{\dagger}
+\frac{1}{2}\Delta_{p}^{\star}(k)c_{-k\alpha}c_{k\alpha}]
\end{eqnarray}
where anisotropic order parameter has the property $\Delta_{p}(-k)=-\Delta_{p}(k)$, and $\alpha$ represents chiral index of the active sub-band.

The embedded, localized site with a single particle level $\epsilon_{i}$ is described by Hamiltonian:
\begin{equation}\label{Hdot_kitaev}
H_{i}=\epsilon_{i}c_{i\alpha}^{\dagger}c_{i\alpha}.
\end{equation}
Lastly, the hybridization Hamiltonian between the localized site and $p$-wave superconductor reads:
\begin{equation}\label{H_hyb}
H_{hyb}^{p}=\sum_{k}[t_{p}c_{k\alpha}^{\dagger}c_{i\alpha}+t_{p}^{\star}c_{i\alpha}^{\dagger}c_{k\alpha}]
\end{equation}

The energy level $\epsilon_{i}$ corresponds to $II$-nd Hubbard sub-gap level $\epsilon^{IIp}_{\downarrow}$, present at each site of the wire from $i=2$ to $N$ as a result of Coulomb repulsion. The recursive summation over these levels, performed in the numerical solution, creates a pair of particle-hole symmetric sub-gap Hubbard bands in the density of states. In Toy Model $I$, the embedded impurity level $\epsilon_{i}$ generates a similar result.

The density of states is calculated from the Green's function matrix of itinerant electron medium (the $\alpha$ index is further suppressed) scattered by the impurity, $\rho(\omega+i\delta)=-(1/\pi)Im[Tr[\hat{g}]]$, with the Hamiltonian $H_{wire}=H_{p}+H_{i}+H_{hyb}^{p}$. The details of calculations can be found in Appendix \ref{toy_model1}.

The calculated density of states  displays two sub-gap resonances, symmetrically in the particle and hole regions, as a result of hybridization of the impurity with superconductor. They are represented in Fig.(\ref{rho TM1}) by two resonances of symmetric shape. In the numerical results, when recursive summation is performed over the sites in the wire, they correspond to Hubbard sub-bands H\textit{h} and H\textit{p} in  Fig.~(\ref{specdensbelow}) and Fig.~(\ref{specdensabove}). For interpretation of the numerical results it is worth to note that the spectrum of the superconductor does not change with replacing $\epsilon_{i}\rightarrow-\epsilon_{i}$, which explains unaltered positions of Hubbard sub-bands H\textit{h/p} when the magnetic field changes.

\textit{Quantum interference between in-gap states.}
The sub-gap particle-hole asymmetric Fano resonances, observed in the density of states of the wire, arise as a result of quantum interference between pairs of local Hubbard levels of the first site $\epsilon_{1\downarrow}^{IIp/h}$ and the pair of the Hubbard sub-bands H\textit{p/h}. We take one pair of the interfering quasiparticle levels: discrete $\epsilon_{1}$ at site $i=1$, and the broad resonance $\epsilon_{0}$ corresponding to the Hubbard sub-band of the rest of the wire.
Our two sub-gap level model can be described by the Hamiltonian:
\begin{equation}\label{two lev quasi}
  H=\sum_{i=0,1}\epsilon_{i}\gamma^{\dagger}_{i}\gamma_{i}+t_{F}(\gamma_{1}^{\dagger}\gamma_{0}+h.c).
\end{equation}

The states $\epsilon_{0}$ and $\epsilon_{1}$ are populated by quasiparticles arising on the onset of $p$-wave superconductivity, when the external magnetic field exceeds the critical magnetic field $V_{z}^{cr}$. The nature of these quasiparticles is revealed by performing Bogoliubov transformation to $p$-wave Hamiltonian Eq.~(\ref{Ham_kitaev}) in a similar way as for $s$-wave superconductor, see for instance \cite{KittelBook}. The obtained quasiparticle operators, which are combinations of particle and hole operators and fulfill fermionic anti-commutation relations ($\alpha$-helical index is suppressed), read as:
\begin{eqnarray}
\label{bogolons}
 \nonumber 
  \gamma_{k}=u_{k}c_{k}-v_{k}c_{-k}^{\dagger} \\
  \gamma_{-k}=u_{k}c_{-k}+v_{k}c_{k}^{\dagger}.
\end{eqnarray}
 The coefficients fulfill the relation $u_{k}^{2}+v_{k}^{2}=1$ and have the values $u_{k}^{2}=(1/2)[1+(\epsilon_{k}/E_{k})]$ and $v_{k}^{2}=(1/2)[1-(\epsilon_{k}/E_{k})]$, where $E_{k}=\sqrt{\epsilon_{k}^{2}+\Delta_{p}^{2}}$.
Adopting these results to the two in-gap levels: the sharp $\epsilon_{1}$, and the broad $\epsilon_{0}$, we define the quasiparticle operators of the levels as $ \gamma_{i}=u_{i}c_{i}-v_{i}c_{i}^{\dagger}$, $(i=0, 1)$, with coefficients $u_{i}=\sqrt{(1/2)[1+(\epsilon_{i}/E_{i})]}$  and $v_{i}=\sqrt{(1/2)[1-(\epsilon_{i}/E_{i})]}$ with $E_{i}=\sqrt{\epsilon_{i}^{2}+\Delta_{p}^{2}}$. The quantities $u_{i}^{2}$ and $v_{i}^{2}$ describe the amount of particles and holes constituting  Bogoliubov quasiparticle, dependent on the position of the level within the gap. For $\epsilon_{i}\gg\epsilon_{F}$  $u_{i}^{2}\rightarrow 1$, for $\epsilon_{i}\ll\epsilon_{F}$  $v_{i}^{2}\rightarrow 1$, and for $\epsilon_{i}=\epsilon_{F}$ $u_{i}^{2}=v_{i}^{2}=1/2$.

The Hamiltonian Eq.~(\ref{two lev quasi}) expressed in terms of creation and annihilation of single particle operators reads:
\begin{eqnarray}\nonumber
H=\sum_{i=0,1}\epsilon_{i}(u_{i}^{2}c_{i}^{\dagger}c_{i}+v_{i}^{2}c_{i}c_{i}^{\dagger})+\\\label{two_lev_quasi2}
t_{F}[C_{1}(c_{0}^{\dagger}c_{1}+h.c)-C_{2}(c_{1}c_{0}+h.c)].
\end{eqnarray}
The tunneling part of the Hamiltonian consists of two terms: the first one with coefficient $C_{1}\equiv u_{1}u_{0}-v_{1}v_{0}$, describing single particle tunneling processes, and the second one with coefficient $C_{2}\equiv v_{1}u_{0}-u_{1}v_{0}$, describing two particle tunneling. These coefficients are dependent on the positions of $\epsilon_{0}$ and $\epsilon_{1}$ in the energy gap.

\begin{figure}
  \centering
  \includegraphics[width=8cm]{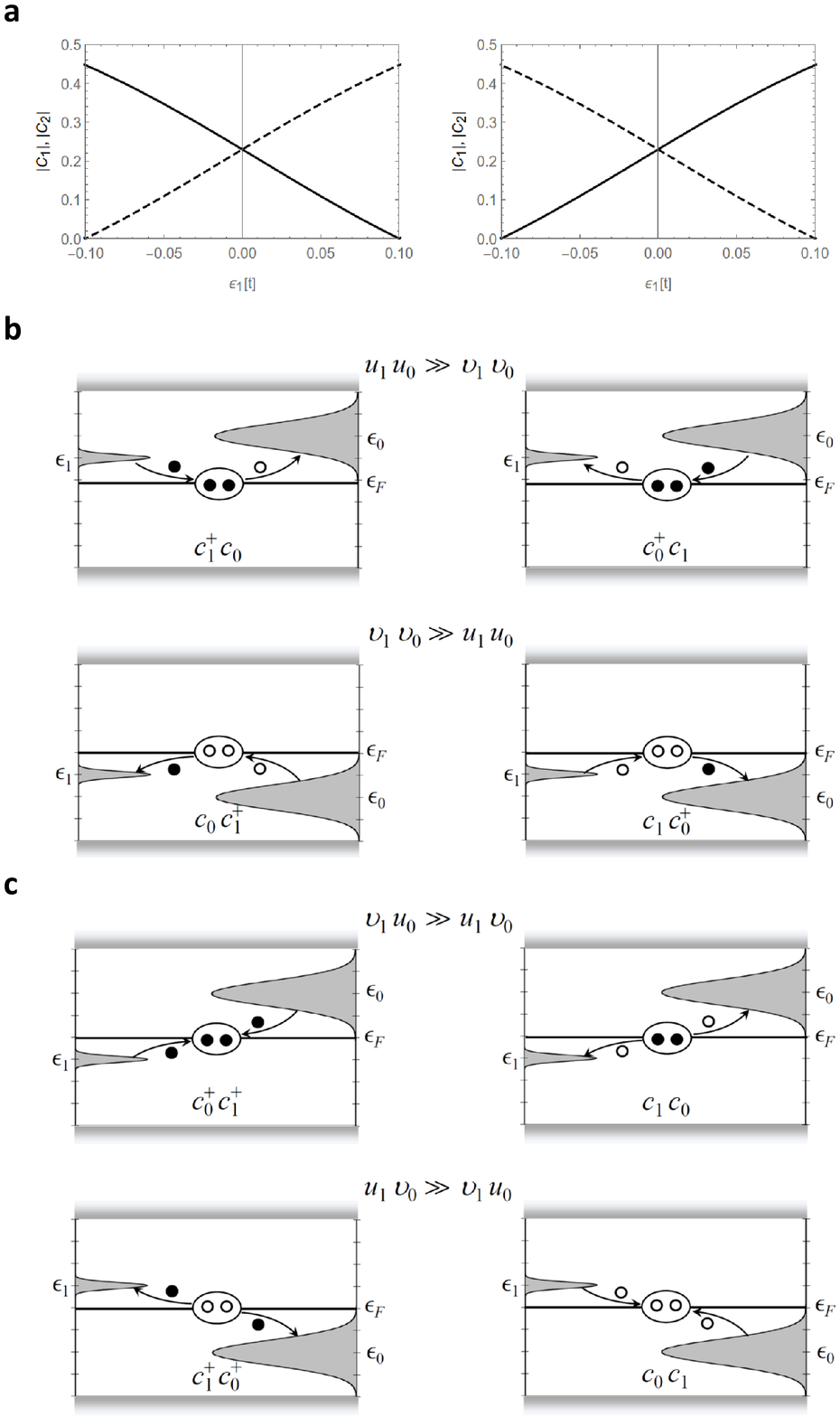}
  \caption{Panel $(a)$: coefficients $|C_{1}|$-dashed line and $|C_{2}|$-solid line dependence on the position of $\epsilon_{1}$ level for the fixed position of $\epsilon_{0}=0.1$ level in the left Panel, and $\epsilon_{0}=-0.1$ in the right Panel. Calculations were performed for $\Delta_{p}=0.2$. Panel $(b)$: schematic of the dominant single-particle tunneling between $\epsilon_{1}$ and $\epsilon_{0}$ levels, when both are positioned in the hole (particle) sector depicted in the upper (lower) part. This arrangement corresponds to magnetic field $V_{z}>V_{z}^{\star}$ . Panel $(c)$: schematic of the dominant two-particle tunneling, when $\epsilon_{1}$ and $\epsilon_{0}$ levels are positioned in different sectors.  This arrangement corresponds to magnetic field $V_{z}<V_{z}^{\star}$.}\label{cum_tun}
\end{figure}

An example of the behavior of $C_{1}$ and $C_{2}$ vs. position of $\epsilon_{1}$ level, for $\epsilon_{0}>\epsilon_{F}$ and $\epsilon_{0}<\epsilon_{F}$ is displayed in Fig.~(\ref{cum_tun}a); schematics of single-particle and two-particle tunneling processes between the $II$-nd Hubbard level of the first site and Hubbard sub-band of the rest of the wire are depicted in panels Fig.~(\ref{cum_tun}b) and Fig.~(\ref{cum_tun}c), respectively. The tunneling takes place via creation of superconducting pairs at Fermi energy.

In Panel $(b)$ of Fig.~(\ref{cum_tun})  processes of single-particle tunneling between the discrete Hubbard quasiparticle level $\epsilon_{1}$ and Hubbard sub-band $\epsilon_{0}$ are depicted. The upper part of the diagram corresponds to  arrangement for $u_{1}u_{0}\gg v_{1}v_{0}$, where both interfering levels are positioned in the particle sector. The left part of the diagram shows the process of tunneling of the particle from $\epsilon_{1}$ quasiparticle level, accompanied by tunneling of the hole in the opposite direction from $\epsilon_{0}$ and, as a result, creation of a  propagating particle-pair at Fermi energy. The right part shows  the process of creation of a particle pair propagating in the opposite direction. The lower part of Panel $(b)$ shows the arrangement for $v_{1}v_{0}\gg u_{1}u_{0}$, where both the interfering levels are positioned in the hole sector. In this case the single-particle tunneling processes effectively create pairs of holes propagating in the opposite direction with respect to the corresponding particle-pairs above.

In Panel $(c)$ of Fig.~(\ref{cum_tun}) processes of two-particle tunneling between corresponding in-gap quasiparticle levels are depicted. Effectively, they create or break apart pairs of particles at Fermi energy.  The upper part of this Panel shows the situation for $v_{1}u_{0}\gg u_{1}v_{0}$ where the discrete Hubbard level $\epsilon_{1}$ is positioned in the hole sector whereas the Hubbard band is situated in the particle sector. In the upper left part, two particles tunneling from $\epsilon_{1}$ and $\epsilon_{0}$ form a pair at Fermi energy, whereas the right part  shows the opposite process of tunneling of holes. The lower part of the Panel $(c)$ describes the processes for the level arrangement $u_{1}v_{0}\gg v_{1}u_{0}$.

In the numerical results there are two particle-hole  symmetric pairs of interfering quasiparticle levels. In our Toy Model it corresponds to the simultaneous interference processes in the upper and the lower part of Panel $(b)$ (Panel $(c)$) for the magnetic field $V_{z}>V_{z}^{\star}$  ($V_{z}< V_{z}^{\star}$). In the upper (lower) part of Panel $(b)$ the $\epsilon_{1}$ quasiparticle level corresponds to the $II$-nd Hubbard level $\epsilon^{IIp}_{1\downarrow}$ ($\epsilon^{IIh}_{1\downarrow}$) and $\epsilon_{0}$ to H\textit{p} (H\textit{h}) Hubbard sub-band. In the upper (lower) part of Panel $(c)$ the $\epsilon_{1}$ quasiparticle level corresponds to the $II$-nd Hubbard level $\epsilon^{IIh}_{1\downarrow}$ ($\epsilon^{IIp}_{1\downarrow}$) and $\epsilon_{0}$ to H\textit{p} (H\textit{h}) Hubbard sub-band.
Also note that the position of $\epsilon_{1}$ has changed in passing from Panel $(b)$ to Panel $(c)$ reflecting its shift by the magnetic field.

From the comparison of Panels $(b)$ and $(c)$ of Fig.~(\ref{cum_tun}) one notes that in the case of two-particle tunneling a process of creation of pairs  at Fermi energy and breaking them apart takes place in the interference process. On the contrary, for the single-particle tunneling it is rather an effective propagation of pairs at Fermi energy with a possible prescribed direction, for instance defined by the direction of propagating particles. This difference is reflected in the Majorana resonance response to those processes: its diminishing by the two-particle processes  as opposed to its robustness to the single-particle tunneling processes.

When both the particle-hole symmetric Hubbard levels $\epsilon^{IIp}_{\downarrow}$ and $\epsilon^{IIh}_{\downarrow}$ are at resonance with Fermi energy, at $V_{z}=V_{z}^{\star}$, the picture of single-particle tunneling changes. It is no longer possible to ascribe the direction of the tunneling pairs; instead creation and breaking apart of the pairs takes place similarly as for two-particle tunneling. As a result, the Majorana resonance is destroyed completely.

It was shown \cite{Stefanski2004d,Miroshnichenko2010}  that Fano resonances in nanoscopic devices can arise as a result of hybridization between $\delta$-like discrete level and a broad level playing the role of continuum of states, present in the original Fano picture \cite{Fano1961}. In such a case the Fano $q$-asymmetry parameter can be determined by the position and the width of the broad level: $q(\omega=0)=-\epsilon_{0}/\Gamma$. This expression results from the mapping of the hybridized two-level system onto the Fano-Anderson model when the continuum of states is replaced by broad localized level. The density of states of the continuum with an embedded impurity, in the Fano-Anderson model, described by the retarded Green's function $g(\omega)$, can be written in terms of the Fano formula \cite{Stefanski2003b}: $\rho(\omega)=\rho_{0}[(\omega+q^{2})/(\omega^{2}+1)]$, where Fano asymmetry parameter $q=-Re g(0)/Im g(0)$. Below, we apply this strategy to our system of two pairs of in-gap states.

Let us calculate the Green's function of the broad $\epsilon_{0}$ quasiparticle level and analyze various scenarios of Fano resonance to appear.
It has a general expression:
\begin{eqnarray}\label{e0bogoliubov}\nonumber
  \langle\langle\gamma_{0}|\gamma_{0}^{\dagger}\rangle\rangle=\langle\langle u_{0}c_{0}-v_{0}c_{0}^{\dagger}|u_{0}c_{0}^{\dagger}-v_{0}c_{0}\rangle\rangle=\\
  u_{0}^{2}\langle\langle c_{0}|c_{0}^{\dagger}\rangle\rangle+v_{0}^{2}\langle\langle c_{0}^{\dagger}|c_{0}\rangle\rangle
  -u_{0}v_{0}[\langle\langle c_{0}|c_{0}\rangle\rangle+\langle\langle c_{0}^{\dagger}|c_{0}^{\dagger}\rangle\rangle].
\end{eqnarray}
Taking into account Hamiltonian (\ref{two_lev_quasi2}), we generate a set of equations of motion for Green's functions required for calculation of  $\langle\langle\gamma_{0}|\gamma_{0}^{\dagger}\rangle\rangle$. The details are shown in Appendix \ref{toy_model1}.

The general equation for $\langle\langle\gamma_{0}|\gamma_{0}^{\dagger}\rangle\rangle$ has a structure too complicated to be listed here, but it has a physically sound form in two limits of interest, related to the numerical results.

Let us start first with the sub-gap level arrangement for $V_{z}>V_{z}^{\star}$, when the interfering pairs of sub-gap states have their positions in the same  particle or hole sector. As we have demonstrated in Fig.~(\ref{cum_tun}a), the single-particle tunneling processes dominate in this arrangement. Thus, assuming $|C_{1}|\gg|C_{2}|$ and setting $C_{2}\equiv0$, we obtain the Green's function:
\begin{equation}\label{gg_Green_C20}
  \langle\langle\gamma_{0}|\gamma_{0}^{\dagger}\rangle\rangle=\frac{u_{0}^{2}}{E_{0-}-\frac{C_{1}^{2}t_{F}^{2}}{E_{1-}+i\delta}+i\Gamma}+
  \frac{v_{0}^{2}}{E_{0+}-\frac{C_{1}^{2}t_{F}^{2}}{E_{1+}+i\delta}+i\Gamma},
\end{equation}
with $E_{i\mp}=\omega\mp\beta_{i}\epsilon_{i}$ and $\beta_{i}=u_{i}^{2}-v_{i}^{2}$ ($i=0, 1$). We have added artificial broadenings $\Gamma$ and $\delta$ ($\Gamma\gg\delta$) of $\epsilon_{0}$ and $\epsilon_{1}$ levels, respectively. Density of states, $\rho_{0}(\omega)=(-1/\pi)Im\langle\langle\gamma_{0}|\gamma_{0}^{\dagger}\rangle\rangle$, following from Eq.~({\ref{gg_Green_C20}) has two-resonance structure with weights $u_{0}^{2}$ and $v_{0}^{2}$ positioned in particle and hole sector, respectively. Indeed, each of the resonances  describes the hybridization of $\epsilon_{0}$ and $\epsilon_{1}$ levels positioned in the same sector. For both levels positioned in the particle sector we can assume that $u_{0}^{2},u_{1}^{2}=1$ and  $v_{0}^{2},v_{1}^{2}=0$, and writing $\epsilon_{0}\equiv \epsilon_{0p}$ and $\epsilon_{1}\equiv \epsilon_{1p}$ we obtain from Eq.~(\ref{gg_Green_C20}):
\begin{equation}\label{gg_e1pe0p}
  \langle\langle\gamma_{0}|\gamma_{0}^{\dagger}\rangle\rangle=\frac{1}{\omega-\epsilon_{0p}-\frac{t_{F}^{2}}{\omega-\epsilon_{1p}+i\delta}+i\Gamma},
\end{equation}
with Fano asymmetry parameter $q_{p}=-\epsilon_{0p}/\Gamma < 0$.

For both interfering levels in the hole sector,  we assume $v_{0}^{2},v_{1}^{2}=1$ and $u_{0}^{2},u_{1}^{2}=0$, as well as $\epsilon_{0}\equiv-\epsilon_{0h}$, $\epsilon_{1}\equiv-\epsilon_{1h}$, and obtain from Eq.~(\ref{gg_Green_C20}):
\begin{equation}\label{gg_e1he0h}
  \langle\langle\gamma_{0}|\gamma_{0}^{\dagger}\rangle\rangle=\frac{1}{\omega+\epsilon_{0h}-\frac{t_{F}^{2}}{\omega+\epsilon_{1h}+i\delta}+i\Gamma},
\end{equation}
with Fano asymmetry parameter $q_{h}=\epsilon_{0h}/\Gamma >0(=-q_{p})$.
Eqs. (\ref{gg_e1pe0p}) and (\ref{gg_e1he0h}) correspond to the Fano resonance curves shown in Panel $(b)$ of Fig.~(\ref{rho TM1}) in the  particle and hole sectors, respectively.

Consider now the sub-gap level arrangement for $V_{z}<V_{z}^{\star}$,  where the interfering pairs of in-gap levels have their positions in different particle and hole sectors. For such an arrangement, see Fig.~(\ref{cum_tun}a), the two-particle tunneling processes dominate. Thus, assuming $|C_{2}|\gg|C_{1}|$ and setting $C_{1}\equiv0$, we obtain the Green's function:
\begin{equation}\label{gg_Green_C10}
  \langle\langle\gamma_{0}|\gamma_{0}^{\dagger}\rangle\rangle=\frac{u_{0}^{2}}{E_{0-}-\frac{C_{2}^{2}t_{F}^{2}}{E_{1+}+i\delta}+i\Gamma}+
  \frac{v_{0}^{2}}{E_{0+}-\frac{C_{2}^{2}t_{F}^{2}}{E_{1-}+i\delta}+i\Gamma},
\end{equation}
which has the structure of two charge-conjugated resonances, each of them describing hybridization of $\epsilon_{0}$ and $\epsilon_{1}$ positioned in different sectors.
For the Hubbard level $\epsilon_{1}\equiv\epsilon_{1p}$ in the particle sector and the broad band $\epsilon_{0}\equiv -\epsilon_{0h}$ in the hole sector we can assume that $u_{0}^{2}, v_{1}^{2}=0$ and $v_{0}^{2}, u_{1}^{2}=1$; thus  Eq.~(\ref{gg_Green_C10}) yields:
\begin{equation}\label{gg_e1pe0h}
  \langle\langle\gamma_{0}|\gamma_{0}^{\dagger}\rangle\rangle=\frac{1}{\omega+\epsilon_{0h}-\frac{t_{F}^{2}}{\omega-\epsilon_{1p}+i\delta}+i\Gamma}.
\end{equation}
with $q_{p}=\epsilon_{0h}/\Gamma>0$.

Simultaneously, for its counterpart: $\epsilon_{1}\equiv-\epsilon_{1h}$ in the hole sector and $\epsilon_{0}\equiv\epsilon_{0p}$ in the particle sector  we assume that $u_{0}^{2}, v_{1}^{2}=1$ and $v_{0}^{2}, u_{1}^{2}=0$ to obtain from Eq.~(\ref{gg_Green_C10}):
\begin{equation}\label{gg_e1he0p}
  \langle\langle\gamma_{0}|\gamma_{0}^{\dagger}\rangle\rangle=\frac{1}{\omega-\epsilon_{0p}-\frac{t_{F}^{2}}{\omega+\epsilon_{1h}+i\delta}+i\Gamma},
\end{equation}
with $q_{h}=-\epsilon_{0p}/\Gamma >0$. Eqs.~(\ref{gg_e1pe0h}) and (\ref{gg_e1he0p}) describe Fano resonances in the particle and the hole sectors, respectively, depicted in Panel $(a)$ of Fig.~(\ref{rho TM1}) with corresponding asymmetry parameters $q_{p}>0$ ad $q_{h}=-q_{p}$.

Regarding the correspondence to the numerical results: as the magnetic field increases, the Fano resonance in the particle region, Eq.~(\ref{gg_e1pe0h}), is shifted into the hole region, described by Eq.~(\ref{gg_e1he0h}), and simultaneously the Fano resonance in the hole region, described by Eq.~(\ref{gg_e1he0p}), is shifted into particle region, described by Eq.~(\ref{gg_e1pe0p}). This exchange in the positions of Fano resonances corresponds to the evolution of the density of states from that depicted in Fig.~(\ref{specdensbelow}) to the one depicted in  Fig.~(\ref{specdensabove}).

\begin{figure}
  \centering
  \includegraphics[width=6cm]{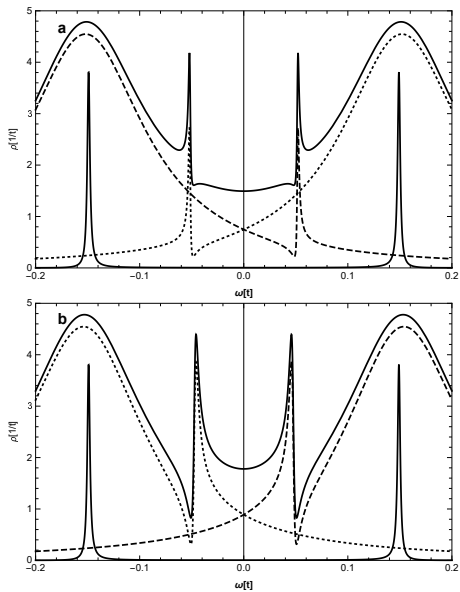}
  \caption{Energy dependence of the density of states and its components obtained within Toy Model $I$. Two symmetric peaks in the particle and hole regions represent the density of states of $p$-wave superconductor with embedded impurity. Panel $(a)$ corresponds to the case of $V_{z}<V_{z}^{\star}$ of the numerical results calculated for the Hubbard resonance in the particle sector $\epsilon_{1}=0.05$, hybridized with $\epsilon_{0}=-0.15$ in the hole sector, and its counterpart Hubbard resonance in the hole sector  $\epsilon_{1}=-0.05$ hybridized with $\epsilon_{0}=0.15$ in the particle sector.   The dashed line corresponds to the density of states with Fano resonance for Hubbard level in the particle sector with $q_{p}=2.14$ and the dotted curve is for Hubbard resonance in the hole region with $q_{h}=-2.14$. The solid curve is the sum of spectral densities for the particle and the hole regions. Panel (b) corresponds to the case of $V_{z}>V_{z}^{\star}$, when the Fano resonances in the particle and the hole sector exchanged their positions in energy scale. As a result, $q_{p}=-2.14$ and $q_{h}=2.14$.  The dependencies are calculated for $\Delta_{p}=0.2$,  $\delta=0.005$, $t_{p}=0.7$, $t_{F}=0.02$, $\Gamma=0.07$ and $\rho_{w}=1/4$.}\label{rho TM1}
\end{figure}

The processes when the two-particle tunneling is dominant and the spectral weight of MZM resonance is diminished and visibly shifted into the Hubbard sub-bands, resemble quasiparticle poisoning of the Majorana peak \cite{Rainis2012a,Karzig2021} by the presence of in-gap states, but in the present case it is realized indirectly, via quantum interference between such states.

As we will show in Toy Model $2$, for $V_{z}=V_{z}^{\star}$, when the discrete, charge-conjugated, Hubbard levels $\epsilon^{IIp}_{1\downarrow}$ and $\epsilon^{IIh}_{1\downarrow}$  match Fermi energy and the Majorana resonance vanishes completely, the first and second order tunneling processes between Majorana and Hubbard levels have equal contribution to quantum interference.

The interference pattern between a pair of in-gap quasiparticle states has its correspondence to the interference process of ionization of an atom into the continuum from its ground state (see Fig. (3) of Miroshnichenko \textit{et.al.} \cite{Miroshnichenko2010}). This process can be realized by a direct ionization of an atom or by  autoionization from its discrete state. Both the processes are quantum mechanically coupled giving rise to Fano resonance in atomic spectrum.  In the present case, the local Hubbard level of the site $i=1$ corresponds to a discrete autoionized state $|d\rangle$,  and the in-gap Hubbard sub-band corresponds to the continuum band $|c\rangle$, both coupled via the superconductor ground state  $|g\rangle$ by the hopping amplitude.

\subsubsection{Toy Model $II$: Majorana bound state coupled to in-gap quasiparticle state }
In Toy Model $II$ we proceed with separating topological superconducting wire into the end site of the wire, $i=1$, with the localized quasiparticle energy level $\epsilon_{1}$, and the rest of the wire. The wire is described here by the simplest Hamiltonian of two hybridized MZM $\lambda_{1}$ and $\lambda_{2}$  at its ends with the strength of  $\epsilon_{m}$.  $\epsilon_{m}$ describes the overlap of the Majorana wave functions, $\epsilon_{m}\sim e^{-L/\xi}$,where $\xi$ is the induced superconducting coherence length and $L$ - the wire length.

The in-gap localized quasiparticle site $\epsilon_{1}$ is coupled to the MZM $\lambda_{1}$ via hopping amplitude $t_{m}$. We are interested in the influence of the localized  $\epsilon_{1}$ site on the density of states of the wire at the Majorana $\lambda_{1}$ site and analysis of the vanishing of the Majorana resonance, obtained in the numerical calculations when $\epsilon^{IIp}_{\downarrow}=\epsilon^{IIh}_{\downarrow}=\epsilon_{F}$.

In the following we suppress the chiral $\alpha$ index of the sector, in which the site-wire hybridization takes place. The in-gap state is described by the quasiparticle operator $\gamma_{1}=u_{1}c_{1}-v_{1}c_{1}^{\dagger}$, with coefficients $u_{1}$ and $v_{1}$ previously defined. The Hamiltonian of our simplified system reads as follows:
\begin{equation}\label{H_2}
H_{2}= \epsilon_{1}\gamma_{1}^{\dagger}\gamma_{1}+t_{m}(\gamma_{1}-\gamma_{1}^{\dagger})\lambda_{1}+i\epsilon_{m}\lambda_{1}\lambda_{2}
\end{equation}

The Majorana operators can be written in terms of fermionic operators: $\lambda_{1}=(f+f^{\dagger})/\sqrt 2$ and $\lambda_{2}=i (f-f^{\dagger})/\sqrt 2$.

Hamiltonian Eq.~(\ref{H_2}), written in the single particle fermionic operators, assumes the form:
\begin{equation}\label{H_2fermion}
  H_{2}=\epsilon_{1}(u_{1}^{2}c_{1}^{\dagger}c_{1}+v_{1}^{2}c_{1}c_{1}^{\dagger})+\tilde{t}_{m}(c_{1}-c_{1}^{\dagger})(f+f^{\dagger})+
  \epsilon_{m}(f^{\dagger}f-\frac{1}{2}),
\end{equation}
where $\tilde{t}_{m}=t_{m}(u_{1}+v_{1})/\sqrt{2}$.
Next we calculate the Green's function of the  Majorana state $\lambda_{1}$  by EOM, utilizing Hamiltonian Eq.~(\ref{H_2fermion}). The EOM method generates the set of equations for Green's functions in $\omega$-domain, which are listed in Appendix \ref{toy_model2}. They yield the solution for Majorana Green's function:
 \begin{equation}\label{MajoranaGFmain}
\langle\langle\lambda_{1}|\lambda_{1}\rangle\rangle=\frac{1}{2}\langle\langle f+f^{\dagger}|f+f^{\dagger}\rangle\rangle=
\frac{\omega}{\omega^2-\epsilon_{m}^2-\frac{2\tilde{t}_{m}^2\omega^2}{\omega^2-\beta_{1}^{2}\epsilon_{1}^2}}
\end{equation}

\begin{figure}
  \centering
  \includegraphics[width=8cm]{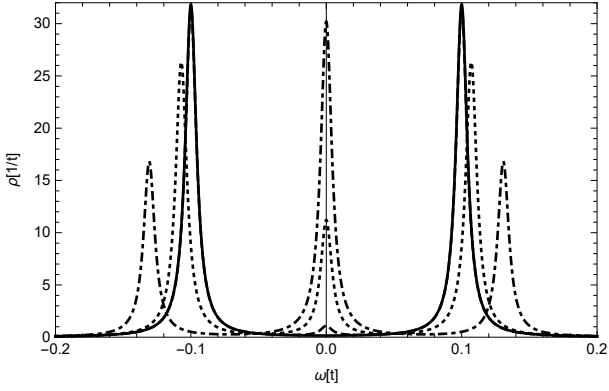}
  \caption{Density of states of the Majorana zero mode coupled to $\epsilon_{1}$ site for various positions of  $\epsilon_{1}$:  $\epsilon_{1}=0.15$- dash-dotted curve, $\epsilon_{1}=0.1$- dotted curve, $\epsilon_{1}=0.05$- dashed curve and $\epsilon_{1}=0$- solid curve. The dependencies were calculated for $\epsilon_{m}=0$, $t_{m}=0.1$, $\Delta_{p}=0.2$ and $\delta=0.005$.}\label{MBS_diminish}
\end{figure}

The MZM selfenergy due to the coupling to $\epsilon_{1}$ site, from Eq.~(\ref{MajoranaGFmain}), is:
\begin{equation}
\Sigma_{1}(\omega)=\frac{2\tilde{t}_{m}^2\omega^2}{\omega^2-\beta_{1}^{2}\epsilon_{1}^2}=\tilde{t}_{m}^2(\frac{1}{\omega-\beta_{1}\epsilon_{1}}+\frac{1}{\omega+\beta_{1}\epsilon_{1}}).
\end{equation}
It has the poles in the particle and the hole regions at $\omega=\pm\beta_{1}\epsilon_{1}$, which in our general model correspond to the pair of Hubbard levels $\epsilon^{IIp}_{\downarrow}$ and $\epsilon^{IIh}_{\downarrow}(=-\epsilon^{IIp}_{\downarrow})$. Shifted by the magnetic field $V_{z}$ towards Fermi energy, they diminish Majorana resonance completely, when in resonance with $\epsilon_{F}$.

The density of states of the Majorana state is calculated from retarded Green's function after performing analytical continuation: $\rho_{MZM}(\omega+i\delta)=-(1/\pi)Im\langle\langle\lambda_{1}|\lambda_{1}\rangle\rangle$.

Evolution of the density of states of the MZM $\lambda_{1}$ for various positions of the coupled $\epsilon_{1}$ level is displayed in Fig.~(\ref{MBS_diminish}). As the discrete quasiparticle level approaches Fermi energy, the central Majorana peak is gradually diminished, and for $\epsilon_{1}=\epsilon_{F}$ disappears completely.  The two particle-hole symmetric resonances in the density of states, which develop at $\omega\simeq\mp\sqrt{\Re\Sigma_{1}(\omega+i\delta)}$, are caused by the coupling of the localized site to a superconductor. They correspond to symmetric resonances reproduced within Toy Model $I$ and Hubbard sub-bands in the general model. For the bare level situated at Fermi energy $\epsilon_{1}=\epsilon_{F}$, they are located exactly at $\omega=\mp t_{m}$.

It is instructive to analyze the influence of the one- and the two-particle tunneling between the quasiparticle state and Majorana state on the Majorana resonance and compare it to the results of tunneling between two sub-gap quasiparticle levels of Toy Model $I$. Let us rewrite the tunneling term in Hamiltonian, Eq.~(\ref{H_2fermion}), and separate one- and two-particle tunneling processes:
\begin{eqnarray}\nonumber
 H_{tun}=\tilde{t}_{m}(c_{1}^{\dagger}f+c_{1}^{\dagger}f^{\dagger}+h.c.)=\\
 \tilde{t}_{m}(c_{1}^{\dagger}f+h.c.)+\tilde{t}_{m}(c_{1}^{\dagger}f^{\dagger}+h.c.).
\end{eqnarray}
  The calculated Majorana Green's function, Eq.~(\ref{MajoranaGFmain}), separately for  one- and two particle processes assumes the same form for $\epsilon_{m}=0$:
\begin{equation}\label{one_tun}
 \langle\langle\lambda_{1}|\lambda_{1}\rangle\rangle^{(1/2)}=\frac{1}{2}\left[\frac{1}{\omega-\frac{\tilde{t}_{m}^{2}}{\omega-\beta_{1}\epsilon_{1}}}+
 \frac{1}{\omega-\frac{\tilde{t}_{m}^{2}}{\omega+\beta_{1}\epsilon_{1}}}\right].
\end{equation}
It demonstrates that in the case of direct tunneling between in-gap quasiparticle state and MZM both tunneling processes have the same contributions to the diminishing of the Majorana resonance.

\subsection{Difference in tunneling amplitude between quasiparticle in-gap state and "accidental" state at Fermi energy compared to Majorana zero mode}
Let us discuss the limiting case of both  quasiparticle levels positioned at Fermi energy. For such an arrangement  $u_{i}=v_{i}=1/\sqrt{2}$ ($i=0,1$), and in Toy Model $I$ the effective hopping between levels $\epsilon_{0}$ and $\epsilon_{1}$ is zero, which can be noticed by the inspection of the Hamiltonian Eq.~(\ref{two_lev_quasi2}). 

There is a substantial difference, however, when one of the Bogoliubov quasiparticle levels is replaced by MZM, as in Toy Model $II$. When the $\epsilon_{1}$ level approaches Fermi energy, the effective hopping $\tilde{t}_{m}$ approaches its maximal value, see Eq.(\ref{H_2fermion}).  This non-zero hopping between MZM and the pair of Hubbard resonances approaching Fermi energy produces complete vanishing of the Majorana resonance, as demonstrated by the numerical results.

The above finding can be related to the recent experimental attempts of distinguishing Majorana zero modes from "accidental" quasiparticle states located at Fermi energy \cite{Frolov2020,Yu2021}.

Tunneling between superconductors possessing in-gap states has been realized experimentally in various configurations \cite{Ruby2015,Huang2020,Rubio-Verdu2021}. For the present discussion to be valid, such tunneling should be realized  between superconductors with non-conserved spin quantum number. Such requirement can be fulfilled for instance in the superconductor hybrid structures with strong synthetic spin-orbit interaction \cite{Lo2014,Desjardins2019}. Suppose that the tunneling current is initiated between two such superconductors, labelled $0$ and $1$. Superconductor $1$ with sub-gap state $\epsilon_{1}$ is coupled to the end site of superconductor $0$. Let us assume for simplicity that the superconducting gaps in both superconductors are comparable in magnitude: $\Delta_{0}\cong\Delta_{1}\equiv\Delta$, but there is a small finite bias $eV$ between them. For simplicity we assume that chemical potential in superconductor $0$ is located at zero energy, $\mu_{0}=0$, and $\mu_{1}$ is shifted by the bias voltage: $\mu_{1}=eV$. Due to the shift of $\mu_{1}$, the particle and the hole coefficients $u_{1}$ and $v_{1}$ of the quasiparticle $\gamma_{1}$ are modified accordingly, and for $\epsilon_{1}=\mu_{1}$ they become $u_{1}=\sqrt{(1/2)[1+(2eV/E_{1})]}$ and $v_{1}=\sqrt{(1/2)[1-(2eV/E_{1})]}$, where $E_{1}=\sqrt{4eV^2+\Delta^2}$.

If an "accidental" quasiparticle state in superconductor $0$ resides at its end and  $\epsilon_{0}=\mu_{0}$, then $u_{0}=v_{0}=1/\sqrt{2}$, and the effective hopping between $\epsilon_{1}$ and $\epsilon_{0}$, following from the Hamitonian Eq.~(\ref{two_lev_quasi2}), is $\tilde{t}_{F}=t_{F}(u_{1}-v_{1})/\sqrt{2}$ with $u_{1}$ and $v_{1}$ dependent on bias voltage. Contrary, when a true MZM appears at zero energy at the end of superconductor $0$, the effective hopping between $\epsilon_{1}$ and MZM, from Eq.~(\ref{H_2fermion}), is $\tilde{t}_{m}=t_{m}(u_{1}+v_{1})/\sqrt{2}$. The tunneling current is governed by the square of the tunneling matrix element. For a small bias $eV \ll \Delta$ we obtain for two coupled Bogoliubov quasiparticle levels $\tilde{t}_{F}^{2}=(t_{F}^{2}/2)(u_{1}-v_{1})^{2} =t_{F}^{2} eV^{2}/\Delta^2$, whereas for the true MZM, there is $\tilde{t}_{m}^{2}=(t_{m}^{2}/2)(u_{1}+v_{1})^{2} =t_{m}^{2}(1- eV^{2}/\Delta^2)$, where in the derivation  the expansion $\sqrt{1-x}\simeq 1-x/2$ is used. Thus, there is a strikingly different dependence on the bias voltage for the "accidental" quasiparticle level at Fermi energy as compared to true MZM. In the first case the tunneling between the levels approaches zero value for a vanishing bias, whereas in the second case it reaches its maximal value.

\section{Concluding remarks}
To summarize, we have shown that on-site Coulomb interactions in $1D$-topological wire exhibit local and global effects in its density of states. Globally, when the $II$-nd Hubbard levels at each site enter the superconducting gap, two particle-hole symmetric Hubbard sub-bands arise in the density of states of the wire. Locally, two discrete Hubbard in-gap states are also visible at each site. Quantum interference between Hubbard sub-bands and discrete in-gap states causes the appearance of Fano resonances in particle and hole sectors. Importantly, for the end-site of the wire, this quantum interference has profound impact on Majorana zero mode, and it depends on the nature of tunneling between discrete Hubbard levels of $i=1$ site and Hubbard sub-bands. We have demonstrated that for two-particle tunneling the Majorana resonance is strongly diminished, whereas one-particle tunneling has negligible influence on it. The nature of the tunneling processes depends on the relative positions of interfering in-gap states and can be tuned by the shift of local states by the magnetic field. For the local particle-hole symmetric Hubbard levels in resonance with Fermi energy, both types of direct tunneling into MZM have the same contribution and Majorana resonance is destroyed completely.

We also discussed the difference in the tunneling amplitude between an in-gap quasiparticle state and an "accidental" state at Fermi energy, compared to the case of tunneling to the true Majorana zero mode. This difference can be utilized for experimental distinction of MZM, when the tunneling between two superconductors with large spin-orbit coupling is investigated.

Finally, we have shown that on-site Coulomb interactions promote topological phase and reduce the value of the critical magnetic field for high charge density.
\appendix

\section{Recursive Green's functions calculations within Hubbard $I$ approximation}\label{AppendixA}

The aim is to calculate Green's function matrix of the $i$-site, written in the Nambu space:
\begin{eqnarray}\label{Gmat}\nonumber
\hat{G}_{i,i}=\left(\begin{array}{c}
c_{i\downarrow}\\
c_{i\uparrow}\\
c_{i\uparrow}^{\dagger}\\
c_{i\downarrow}^{\dagger}
\end{array} \right)\otimes\left(c_{i\downarrow}^{\dagger},  c_{i\uparrow}^{\dagger}, c_{i\uparrow}, c_{i\downarrow}\right)=\\
\left(
\begin{array}{cccc}
\langle\langle c_{i\downarrow}|c_{i\downarrow}^{\dagger}\rangle\rangle & \langle\langle c_{i\downarrow}|c_{i\uparrow}^{\dagger}\rangle\rangle &
\langle\langle c_{i\downarrow}|c_{i\uparrow}\rangle\rangle & \langle\langle c_{i\downarrow}|c_{i\downarrow}\rangle\rangle\\
\langle\langle c_{i\uparrow}|c_{i\downarrow}^{\dagger}\rangle\rangle & \langle\langle c_{i\uparrow}|c_{i\uparrow}^{\dagger}\rangle\rangle &
\langle\langle c_{i\uparrow}|c_{i\uparrow}\rangle\rangle & \langle\langle c_{i\uparrow}|c_{i\downarrow}\rangle\rangle\\
\langle\langle c_{i\uparrow}^{\dagger}|c_{i\downarrow}^{\dagger}\rangle\rangle & \langle\langle c_{i\uparrow}^{\dagger}|c_{i\uparrow}^{\dagger}\rangle\rangle &
\langle\langle c_{i\uparrow}^{\dagger}|c_{i\uparrow}\rangle\rangle & \langle\langle c_{i\uparrow}^{\dagger}|c_{i\downarrow}\rangle\rangle\\
\langle\langle c_{i\downarrow}^{\dagger}|c_{i\downarrow}^{\dagger}\rangle\rangle & \langle\langle c_{i\downarrow}^{\dagger}|c_{i\uparrow}^{\dagger}\rangle\rangle &
\langle\langle c_{i\downarrow}^{\dagger}|c_{i\uparrow}\rangle\rangle & \langle\langle c_{i\downarrow}^{\dagger}|c_{i\downarrow}\rangle\rangle
\end{array}
\right)
\end{eqnarray}
Each of the matrix elements of $\hat{G}_{i,j}$ is calculated by equation of motion (EOM) method. On-site Coulomb interaction has been treated within the Hubbard $I$ approximation, in which spin-flip processes are neglected. Density of states of a localized level in this approximation displays two Hubbard resonances at $\epsilon_{\sigma}$ and $\epsilon_{\sigma}+U$ with spectral weights $(1-\langle n_{\bar{\sigma}}\rangle)$ and $\langle n_{\bar{\sigma}}\rangle$, respectively. Subjected to superconducting environment they become quasiparticle levels and acquire their charge-conjugated partners.

A list of EOMs for Green's functions in $\omega$-domain with local interactions $U$ and $\Delta$ is presented below:

\begin{eqnarray}
  (\omega-\epsilon_{\downarrow})\langle\langle c_{i\downarrow}| c_{i\downarrow}^{\dagger}\rangle\rangle=1-\Delta\langle\langle c_{i\uparrow}^{\dagger}| c_{i\downarrow}^{\dagger}\rangle\rangle+U\langle\langle n_{i\uparrow}c_{i\downarrow}| c_{i\downarrow}^{\dagger}\rangle\rangle \\
  (\omega-\epsilon_{\uparrow})\langle\langle c_{i\uparrow}| c_{i\uparrow}^{\dagger}\rangle\rangle=1+\Delta\langle\langle c_{i\downarrow}^{\dagger}| c_{i\uparrow}^{\dagger}\rangle\rangle+U\langle\langle n_{i\downarrow}c_{i\uparrow}| c_{i\uparrow}^{\dagger}\rangle\rangle \\
  (\omega+\epsilon_{\downarrow})\langle\langle c_{i\downarrow}^{\dagger}| c_{i\downarrow}\rangle\rangle=1+\Delta^{\star}\langle\langle c_{i\uparrow}| c_{i\downarrow}\rangle\rangle-U\langle\langle n_{i\uparrow}c_{i\downarrow}^{\dagger}| c_{i\downarrow}\rangle\rangle \\
  (\omega+\epsilon_{\uparrow})\langle\langle c_{i\uparrow}^{\dagger}| c_{i\uparrow}\rangle\rangle=1-\Delta^{\star}\langle\langle c_{i\downarrow}| c_{i\uparrow}\rangle\rangle-U\langle\langle n_{i\downarrow}c_{i\uparrow}^{\dagger}| c_{i\uparrow}\rangle\rangle.
\end{eqnarray}
At this stage the above equations are exact. The equations for Green's functions non-diagonal in spin indices can be easily generated from the above set of equations. In the next step we perform Hubbard $I$ approximation for Green's functions describing in-site Coulomb interactions:
\begin{widetext}
\begin{eqnarray}
(\omega-\epsilon_{\downarrow}-U)\langle\langle n_{i\uparrow}c_{i\downarrow}| c_{i\downarrow}^{\dagger}\rangle\rangle=\langle n_{i\uparrow}\rangle-\Delta\langle\langle c_{i\uparrow}^{\dagger}| c_{i\downarrow}^{\dagger}\rangle\rangle+\Delta\langle\langle n_{i\downarrow}c_{i\uparrow}^{\dagger}| c_{i\downarrow}^{\dagger}\rangle\rangle\\
(\omega-\epsilon_{\uparrow}-U)\langle\langle n_{i\downarrow}c_{i\uparrow}| c_{i\uparrow}^{\dagger}\rangle\rangle=\langle n_{i\downarrow}\rangle+\Delta\langle\langle c_{i\downarrow}^{\dagger}| c_{i\uparrow}^{\dagger}\rangle\rangle-\Delta\langle\langle n_{i\uparrow}c_{i\downarrow}^{\dagger}| c_{i\uparrow}^{\dagger}\rangle\rangle\\
(\omega+\epsilon_{\downarrow}+U)\langle\langle n_{i\uparrow}c_{i\downarrow}^{\dagger}| c_{i\downarrow}\rangle\rangle=\langle n_{i\uparrow}\rangle+\Delta^{\star}\langle\langle c_{i\uparrow}| c_{i\downarrow}\rangle\rangle-\Delta^{\star}\langle\langle n_{i\downarrow}c_{i\uparrow}| c_{i\uparrow}\rangle\rangle\\
(\omega+\epsilon_{\uparrow}+U)\langle\langle n_{i\downarrow}c_{i\uparrow}^{\dagger}| c_{i\uparrow}\rangle\rangle=\langle n_{i\downarrow}\rangle-\Delta^{\star}\langle\langle c_{i\downarrow}| c_{i\uparrow}\rangle\rangle+\Delta^{\star}\langle\langle n_{i\uparrow}c_{i\downarrow}| c_{i\uparrow}\rangle\rangle\\[0.3cm]
(\omega-\epsilon_{\downarrow}-U)\langle\langle n_{i\uparrow}c_{i\downarrow}| c_{i\uparrow}\rangle\rangle=\langle c_{i\uparrow}c_{i\downarrow}\rangle-\Delta\langle\langle c_{i\uparrow}^{\dagger}| c_{i\uparrow}\rangle\rangle+\Delta\langle\langle n_{i\downarrow}c_{i\uparrow}^{\dagger}| c_{i\uparrow}\rangle\rangle\\
(\omega-\epsilon_{\uparrow}-U)\langle\langle n_{i\downarrow}c_{i\uparrow}| c_{i\downarrow}\rangle\rangle=\langle c_{i\downarrow}c_{i\uparrow}\rangle+\Delta\langle\langle c_{i\downarrow}^{\dagger}| c_{i\downarrow}\rangle\rangle-\Delta\langle\langle n_{i\uparrow}c_{i\downarrow}^{\dagger}| c_{i\downarrow}\rangle\rangle\\
(\omega+\epsilon_{\downarrow}+U)\langle\langle n_{i\uparrow}c_{i\downarrow}^{\dagger}| c_{i\uparrow}^{\dagger}\rangle\rangle=\langle c_{i\downarrow}^{\dagger}c_{i\uparrow}^{\dagger}\rangle+\Delta^{\star}\langle\langle c_{i\uparrow}| c_{i\uparrow}^{\dagger}\rangle\rangle-\Delta^{\star}\langle\langle n_{i\downarrow}c_{i\uparrow}| c_{i\uparrow}^{\dagger}\rangle\rangle\\
(\omega+\epsilon_{\uparrow}+U)\langle\langle n_{i\downarrow}c_{i\uparrow}^{\dagger}| c_{i\downarrow}^{\dagger}\rangle\rangle=\langle c_{i\uparrow}^{\dagger}c_{i\downarrow}^{\dagger}\rangle-\Delta^{\star}\langle\langle c_{i\downarrow}| c_{i\downarrow}^{\dagger}\rangle\rangle+\Delta^{\star}\langle\langle n_{i\uparrow}c_{i\downarrow}| c_{i\downarrow}^{\dagger}\rangle\rangle
\end{eqnarray}
\end{widetext}
Taking into account the above set of equations, it is convenient to write the Dyson equation for the local Green's function as a matrix:
\begin{equation}\label{g_loc}
\hat{g}_{loc}=\left[[\hat{g}_{0}]^{-1}-\hat{V} \right]^{-1}
\end{equation}
with matrices
\begin{widetext}
\begin{eqnarray}\label{g0int}
\hat{g}_{0}=\left(\begin{array}{cccc}
\frac{\omega-\epsilon_{\downarrow}-U(1-\langle n_{i,\uparrow}\rangle)}{(\omega-\epsilon_{\downarrow})(\omega-\epsilon_{\downarrow}-U)} &
-\frac{U\langle c^{\dagger}_{i,\uparrow}c_{i,\downarrow}\rangle}{(\omega-\epsilon_{\downarrow})(\omega-\epsilon_{\downarrow}-U)} &
\frac{U\langle c_{i,\uparrow}c_{i,\downarrow}\rangle}{(\omega-\epsilon_{\downarrow})(\omega-\epsilon_{\downarrow}-U)} & 0\\
-\frac{U\langle c^{\dagger}_{i,\downarrow}c_{i,\uparrow}\rangle}{(\omega-\epsilon_{\uparrow})(\omega-\epsilon_{\uparrow}-U)} &
\frac{\omega-\epsilon_{\uparrow}-U(1-\langle n_{i,\downarrow}\rangle)}{(\omega-\epsilon_{\uparrow})(\omega-\epsilon_{\uparrow}-U)} & 0 &
\frac{U\langle c_{i,\downarrow}c_{i,\uparrow}\rangle}{(\omega-\epsilon_{\uparrow})(\omega-\epsilon_{\uparrow}-U)}\\
\frac{U\langle c_{i,\downarrow}^{\dagger}c_{i,\uparrow}^{\dagger}\rangle}{(\omega+\epsilon_{\uparrow})(\omega+\epsilon_{\uparrow}+U)} & 0 &
\frac{\omega+\epsilon_{\uparrow}+U(1-\langle n_{i,\downarrow}\rangle)}{(\omega+\epsilon_{\uparrow})(\omega+\epsilon_{\uparrow}+U)} &
\frac{-U\langle c_{i,\downarrow}c_{i,\uparrow}^{\dagger}\rangle}{(\omega+\epsilon_{\uparrow})(\omega+\epsilon_{\uparrow}+U)}\\
0 & \frac{U\langle c_{i,\uparrow}^{\dagger}c_{i,\downarrow}^{\dagger}\rangle}{(\omega+\epsilon_{\downarrow})(\omega+\epsilon_{\downarrow}+U)} &
\frac{-U\langle c_{i,\uparrow}c_{i,\downarrow}^{\dagger}\rangle}{(\omega+\epsilon_{\downarrow})(\omega+\epsilon_{\downarrow}+U)} &
\frac{\omega+\epsilon_{\downarrow}+U(1-\langle n_{i,\uparrow}\rangle)}{(\omega+\epsilon_{\downarrow})(\omega+\epsilon_{\downarrow}+U)},
\end{array}
\right)
\end{eqnarray}
\end{widetext}
and
\begin{eqnarray}
\hat{V}=\left(
\begin{array}{cccc}
0 & 0 & -\Delta & 0\\
0 & 0 & 0 & \Delta\\
-\Delta & 0 & 0 & 0\\
0 & \Delta & 0 & 0
\end{array}
\right),
\end{eqnarray}
where the notation is used: $\epsilon_{\downarrow/\uparrow}=-\mu\mp V_{z}$. In the numerical calculations the analytic continuation $\omega\rightarrow\omega+i\delta$ has been performed. Numerically the local Green's function matrix is calculated for a set of input values of correlators, which are then found selfconsistently.

Let us analyze non-local contributions to the Green's function matrix due to tight-binding and Rashba hoppings. The set of EOMs for diagonal in spin indices particle and hole Green's functions generated by these hopping reads as follows:
\begin{widetext}
\begin{eqnarray}
(\omega-\epsilon_{\downarrow})\langle\langle c_{i\downarrow}|c_{i\downarrow}^{\dagger}\rangle\rangle=1-t\langle\langle c_{i-1\downarrow}|c_{i\downarrow}^{\dagger}\rangle\rangle-t\langle\langle c_{i+1\downarrow}|c_{i\downarrow}^{\dagger}\rangle\rangle-t_{so}\langle\langle c_{i-1\uparrow}|c_{i\downarrow}^{\dagger}\rangle\rangle+t_{so}\langle\langle c_{i+1\uparrow}|c_{i\downarrow}^{\dagger}\rangle\rangle\\
(\omega-\epsilon_{\uparrow})\langle\langle c_{i\uparrow}|c_{i\uparrow}^{\dagger}\rangle\rangle=1-t\langle\langle c_{i-1\uparrow}|c_{i\uparrow}^{\dagger}\rangle\rangle-t\langle\langle c_{i+1\uparrow}|c_{i\uparrow}^{\dagger}\rangle\rangle+t_{so}\langle\langle c_{i-1\downarrow}|c_{i\uparrow}^{\dagger}\rangle\rangle-t_{so}\langle\langle c_{i+1\downarrow}|c_{i\uparrow}^{\dagger}\rangle\rangle\\
(\omega+\epsilon_{\downarrow})\langle\langle c_{i\downarrow}^{\dagger}|c_{i\downarrow}\rangle\rangle=1+t\langle\langle c_{i-1\downarrow}^{\dagger}|c_{i\downarrow}\rangle\rangle+t\langle\langle c_{i+1\downarrow}^{\dagger}|c_{i\downarrow}\rangle\rangle+t_{so}\langle\langle c_{i-1\uparrow}^{\dagger}|c_{i\downarrow}\rangle\rangle-t_{so}\langle\langle c_{i+1\uparrow}^{\dagger}|c_{i\downarrow}\rangle\rangle\\
(\omega+\epsilon_{\uparrow})\langle\langle c_{i\uparrow}^{\dagger}|c_{i\uparrow}\rangle\rangle=1+t\langle\langle c_{i-1\uparrow}^{\dagger}|c_{i\uparrow}\rangle\rangle+t\langle\langle c_{i+1\uparrow}^{\dagger}|c_{i\uparrow}\rangle\rangle-t_{so}\langle\langle c_{i-1\downarrow}^{\dagger}|c_{i\uparrow}\rangle\rangle+t_{so}\langle\langle c_{i+1\downarrow}^{\dagger}|c_{i\uparrow}\rangle\rangle
\end{eqnarray}
\end{widetext}
From the form of these equations, it is convenient to define hopping matrices:
\begin{eqnarray}
\hat{t}=\left(
\begin{array}{cccc}
-t & 0 & 0 & 0\\
0 & -t & 0 & 0\\
0 & 0 & t & 0\\
0 & 0 & 0 & t
\end{array}
\right),\\
 \hat{t}^{so}_{L}=\left(
\begin{array}{cccc}
0 & t_{so} & 0 & 0\\
-t_{so} & 0 & 0 & 0\\
0 & 0 & 0 & t_{so}\\
0 & 0 & -t_{so} & 0
\end{array}
\right)\\
\hat{t}^{so}_{R}=-\hat{t}^{so}_{L}\\
\hat{t}_{L}=\hat{t}+\hat{t}^{so}_{L} \\
\hat{t}_{R}=\hat{t}+\hat{t}^{so}_{R},
\end{eqnarray}
where the subscript $L$ ($R$) describes the direction of propagation inside the wire.
In the next step the Dyson equation for the Green's function matrix, Eq.~(\ref{Gmat}), is formulated and calculated recursively \cite{Lee1981,MacKinnon1985,Asano2001,Potter2011}, taking into account all the sites present in the wire:
\begin{eqnarray}\label{last}
\hat{G}_{i,i}=\hat{g}_{i,i}+\hat{g}_{i,i}\hat{V}\hat{G}_{i,i}+\hat{g}_{i,i}\hat{t}_{L}\hat{G}_{i-1,i}\\\label{first}
\hat{G}_{i,i}=\hat{g}_{i,i}+\hat{g}_{i,i}\hat{V}\hat{G}_{i,i}+\hat{g}_{i,i}\hat{t}_{R}\hat{G}_{i+1,i}\\\label{middle}
\hat{G}_{i,i}=\hat{g}_{i,i}+\hat{g}_{i,i}\hat{V}\hat{G}_{i,i}+\hat{g}_{i,i}\hat{t}_{L}\hat{G}_{i-1,i}+\hat{g}_{i,i}\hat{t}_{R}\hat{G}_{i+1,i}.
\end{eqnarray}
Eqs.(\ref{last}), (\ref{first}) and (\ref{middle}) describe recursive summations for the last site, first site and any other site in the wire, respectively.
Taking into account relations between the subsequent sites:
\begin{eqnarray}
\hat{G}_{i-1,i}=\hat{g}_{i,i}\hat{t}_{L}^{\dagger}\hat{G}_{i,i}\\
\hat{G}_{i+1,i}=\hat{g}_{i,i}\hat{t}_{R}^{\dagger}\hat{G}_{i,i},
\end{eqnarray}
the general recursive expression is of the form:
\begin{eqnarray}\nonumber
\hat{G}_{i,i}=
\hat{g}_{i,i}+\hat{g}_{i,i}\hat{V}\hat{G}_{i,i}+\hat{g}_{i,i}\hat{t}_{L}\hat{g}_{i,i}\hat{t}_{L}^{\dagger}\hat{G}_{i,i}\\
+\hat{g}_{i,i}\hat{t}_{L}\hat{g}_{i,i}\hat{t}_{R}^{\dagger}\hat{G}_{i,i}
\end{eqnarray}

The recurrence calculation of the Green's function matrix for a given site is performed in two steps. Firstly, the local Green's function, Eq~(\ref{g_loc}), is calculated for each $\omega$ value with a given set of input correlator values. Then the recurrence is performed with the first matrix input $\hat{G}_{in}=\hat{g}_{loc}$:
\begin{equation}
\hat{G}=\left[[\hat{g}_{loc}]^{-1}-\hat{t}_{R}\hat{G}_{in}\hat{t}_{R}^{\dagger}-\hat{t}_{L}\hat{G}_{in}\hat{t}_{L}^{\dagger} \right]^{-1}.
\end{equation}
When recursive summation is completed, the correlators are again calculated and compared to those calculated in the previous step, checking if the selfconsistency condition is met. If it is fulfilled, the loop is terminated and the density of states is calculated.  For the topological state there are six independent correlators to be found: occupancies $\langle n_{\downarrow}\rangle$ and $\langle n_{\uparrow}\rangle$, $s$-wave correlators $\langle c_{\downarrow}c_{\uparrow}\rangle$ and $\langle c_{\uparrow}c_{\downarrow}\rangle$, and Rashba correlators $\langle c_{\downarrow}c_{\uparrow}^{\dagger}\rangle$ and $\langle c_{\downarrow}^{\dagger}c_{\uparrow}\rangle$. The remaining correlators are found from the relations $\langle c_{\downarrow}^{\dagger}c_{\uparrow}^{\dagger}\rangle=\langle c_{\uparrow}c_{\downarrow}\rangle$, $\langle c_{\uparrow}^{\dagger}c_{\downarrow}^{\dagger}\rangle=\langle c_{\downarrow}c_{\uparrow}\rangle$, and $\langle c_{\uparrow}c_{\downarrow}^{\dagger}\rangle=\langle c_{\downarrow}c_{\uparrow}^{\dagger}\rangle$, $\langle c_{\uparrow}^{\dagger}c_{\downarrow}\rangle=\langle c_{\downarrow}^{\dagger}c_{\uparrow}\rangle$. These relations follow from the relation between retarded Green's functions: $[\hat{G}_{i,i}(\omega)]_{k,l}=[\hat{G}_{i,i}(-\omega)]_{l,k}^{*}$, valid for zero magnetic field. Despite the magnetic field $V_{z}$ is non-zero in the calculations and it initiates topological phase, this new emergent phase effectively involves no magnetic field.   At each selfconsistency step the correlators are calculated from the corresponding matrix elements of the Green's function matrix, Eq.~(\ref{Gmat}), obtained by recursive summation:
\begin{eqnarray}
\langle n_{i\downarrow}\rangle=-\frac{1}{\pi}\int_{-\infty}^{0}d\omega Im[\hat{G}_{i,i}(\omega)]_{1,1},\\
\langle n_{i\uparrow}\rangle=-\frac{1}{\pi}\int_{-\infty}^{0}d\omega Im[\hat{G}_{i,i}(\omega)]_{2,2},\\
\langle c_{i\downarrow}c_{i\uparrow}\rangle=-\frac{1}{\pi}\int_{-\infty}^{0}d\omega Im[\hat{G}_{i,i}(\omega)]_{2,4},\\
\langle c_{i\uparrow}c_{i\downarrow}\rangle=-\frac{1}{\pi}\int_{-\infty}^{0}d\omega Im[\hat{G}_{i,i}(\omega)]_{1,3},\\
\langle c_{i\downarrow}c_{i\uparrow}^{\dagger}\rangle=-\frac{1}{\pi}\int_{-\infty}^{0}d\omega Im[\hat{G}_{i,i}(\omega)]_{3,4},\\
\langle c_{i\downarrow}^{\dagger}c_{i\uparrow}\rangle=-\frac{1}{\pi}\int_{-\infty}^{0}d\omega Im[\hat{G}_{i,i}(\omega)]_{2,1}.
\end{eqnarray}
Finally, the density of states $\rho_(\omega)=\sum_{\sigma=\downarrow,\uparrow}\rho_{\sigma}(\omega)=-(1/\pi)Im[[\hat{G}_{1,1}(\omega)]_{1,1}+[\hat{G}_{1,1}(\omega)]_{2,2}]$ is calculated with the determined values of the correlators.

\section{Toy Model $I$: calculation details}\label{toy_model1}
\subsection{Topological superconductor with an embedded impurity}
The general Green's function matrix of superconducting electron medium ($\alpha$-index is suppressed) is of the form:
\begin{eqnarray}\nonumber
\hat{g}=\sum_{k}\left(\begin{array}{c}
c_{k}\\
c_{-k}^{\dagger}
\end{array} \right)\otimes\left(c^{\dagger}_{k},  c_{-k}\right)=\\
\sum_{k}\left(
\begin{array}{cc}
\langle\langle c_{k}|c_{k}^{\dagger}\rangle\rangle & \langle\langle c_{k}|c_{-k}\rangle\rangle\\
\langle\langle c_{-k}^{\dagger}|c_{k}^{\dagger}\rangle\rangle & \langle\langle c_{-k}^{\dagger}|c_{-k}\rangle\rangle
\end{array}
\right)
\end{eqnarray}
To do so, it is convenient \cite{Hewson1993}  to write the  matrix of Dyson equation, $\hat{g}={[\hat{g}_{0}]^{-1}-\hat{\Sigma}}^{-1}$, in the form of $\hat{T}$-matrix, $\hat{g}=\hat{g}_{0}+\hat{g}_{0}\hat{T}\hat{g}_{0}$, where the $\hat{T}$-matrix is expressed in terms of selfenergy $\hat{\Sigma}$: $\hat{T}=\hat{\Sigma}(\hat{1}-\hat{g}_{0}\hat{\Sigma})^{-1}$. The localized state described by Green's function matrix $\hat{G}_{i}$ plays the role of scatterer: $\hat{g}=\hat{g}_{0}+\hat{g}_{0}\hat{t}\hat{G}_{i}\hat{t}^{\star}\hat{g}_{0}$. The hopping matrix $\hat{t}$ is diagonal, with matrix elements $t_{p}$.

Dyson equation matrix for $p$-wave superconductor bare Green's function reads:
\begin{equation}\label{dyson_g00}
\hat{g}_{0}=[[\hat{g}^{0}_{0}]^{-1}-\hat{V}]^{-1},
\end{equation}
where:
\begin{equation}
\hat{g}^{0}_{0}=\sum_{k}\left(
\begin{array}{cc}
\frac{1}{\omega-\epsilon_{k}} & 0\\
0 & \frac{1}{\omega+\epsilon_{-k}}
\end{array}
\right)
\end{equation}
and
\begin{equation}
\hat{V}=\left(
\begin{array}{cc}
0 & \Delta^{\star}_{p}\\
\Delta_{p} & 0
\end{array}
\right).
\end{equation}

This gives the  matrix of $p$-wave superconductor (it is assumed for dispersion relations for particles and holes $\epsilon_{k}=\epsilon_{-k}$:
\begin{eqnarray}
\hat{g}_{0}=\sum_{k}\frac{1}{D_{p}}\left(
\begin{array}{cc}
\omega+\epsilon_{-k} & \Delta_{p}\\
\Delta_{p} & \omega-\epsilon_{k}
\end{array}
\right)\\
D_{p}=(\omega+\epsilon_{-k})(\omega-\epsilon_{k})-\Delta_{p}^{2}=\omega^{2}-E_{k}^{2}\\
E_{k}=\sqrt{\epsilon_{k}^{2}+\Delta_{p}^{2}}
\end{eqnarray}
After performing $k$-summation within the gap we obtain:
\begin{equation}
\hat{g}_{0}=\frac{\pi\rho_{0}}{\sqrt{\omega^{2}-\Delta_{p}^{2}}}\left(
\begin{array}{cc}
-\omega & \Delta_{p}\\
\Delta_{p} & -\omega
\end{array}
\right)
\end{equation}
Green's function matrix of the localized site embedded in the superconducting medium, written in Nambu space, has the form:
\begin{eqnarray}\label{Gd_Nambu}
\hat{G}_{i}=\left(\begin{array}{c}
c_{i}\\
c_{i}^{\dagger}
\end{array} \right)\otimes\left(c_{i}^{\dagger},  c_{i}\right)=\left(
\begin{array}{cc}
\langle\langle c_{i}|c_{i}^{\dagger}\rangle\rangle & \langle\langle c_{i}|c_{i}\rangle\rangle\\
\langle\langle c_{i}^{\dagger}|c_{i}^{\dagger}\rangle\rangle & \langle\langle c_{i}^{\dagger}|c_{i}\rangle\rangle
\end{array}
\right).
\end{eqnarray}
Its matrix elements can be calculated within EOM method from the Hamiltonian $H=H_{p}+H_{i}+H^{p}_{hyb}$, Eqs.~(\ref{Ham_kitaev})-(\ref{H_hyb}):
\begin{eqnarray}\nonumber
\hat{G}_{i}=\\
\frac{1}{D}\left(
\begin{array}{cc}
\omega+\epsilon_{i}-\sum_{k}\frac{t_{p}^{2}(\omega-\epsilon_{-k})}{\omega^{2}-E_{k}^{2}} & \Delta_{p}\sum_{k}\frac{t_{p}^{2}}{\omega^{2}-E_{k}^{2}}\\
\Delta_{p}\sum_{k}\frac{t_{p}^{2}}{\omega^{2}-E_{k}^{2}} & \omega-\epsilon_{i}-\sum_{k}\frac{t_{p}^{2}(\omega+\epsilon_{-k})}{\omega^{2}-E_{k}^{2}}
\end{array}
\right)
\end{eqnarray}
where $D$ is the determinant of the above matrix.
After performing summation over $k$ in the sub-gap regime; $\sum_{k}\rightarrow\rho_{0}\int_{-\Delta_{p}}^{\Delta_{p}}d\epsilon$, we obtain:
\begin{eqnarray}
\hat{G}_{i}=\frac{1}{D}\left(
\begin{array}{cc}
\omega+\epsilon_{i}+\frac{\Gamma_{p}\omega}{\sqrt{\Delta_{p}^{2}-\omega^{2}}} & \frac{\Delta_{p}\Gamma_{p}}{\sqrt{\Delta_{p}^{2}-\omega^{2}}}\\
\frac{\Delta_{p}\Gamma_{p}}{\sqrt{\Delta_{p}^{2}-\omega^{2}}} & \omega-\epsilon_{i}+\frac{\Gamma_{p}\omega}{\sqrt{\Delta_{p}^{2}-\omega^{2}}}
\end{array}
\right)
\end{eqnarray}
where $\Gamma_{p}=\pi t_{p}^{2}\rho_{w}$. In the model calculations the density of states in the wire has been assumed to be constant and equal: $\rho_{w}=1/(4t)$.

The location of the in-gap states is determined from the poles of the $\hat{T}$-matrix. As we are interested in the sub-gap regime, where electrons enter the gap only virtually, these states are represented by Dirac delta peaks with infinite lifetime; for numerical calculations an artificial broadening has been introduced.

\subsection{Quantum interference between the Hubbard sub-band and the discrete Hubbard level}
Taking into account Hamiltonian (\ref{two_lev_quasi2}), we generate a set of equations of motion for Green's functions required for calculation of  $\langle\langle\gamma_{0}|\gamma_{0}^{\dagger}\rangle\rangle$. These are as follows:
\begin{eqnarray}
  E_{0-}\langle\langle c_{0}|c_{0}^{\dagger}\rangle\rangle=1+t_{F}C_{1}\langle\langle c_{1}|c_{0}^{\dagger}\rangle\rangle-t_{F}C_{2}\langle\langle c_{1}^{\dagger}|c_{0}^{\dagger}\rangle\rangle \\
   E_{0+}\langle\langle c_{0}^{\dagger}|c_{0}^{\dagger}\rangle\rangle=-t_{F}C_{1}\langle\langle c_{1}^{\dagger}|c_{0}^{\dagger}\rangle\rangle+t_{F}C_{2}\langle\langle c_{1}|c_{0}^{\dagger}\rangle\rangle \\
   E_{0-}\langle\langle c_{0}|c_{0}\rangle\rangle=t_{F}C_{1}\langle\langle c_{1}|c_{0}\rangle\rangle-t_{F}C_{2}\langle\langle c_{1}^{\dagger}|c_{0}\rangle\rangle \\
   E_{0+}\langle\langle c_{0}^{\dagger}|c_{0}\rangle\rangle=1-t_{F}C_{1}\langle\langle c_{1}^{\dagger}|c_{0}\rangle\rangle+t_{F}C_{2}\langle\langle c_{1}|c_{0}\rangle\rangle,
\end{eqnarray}
and
\begin{eqnarray}
  E_{1-}\langle\langle c_{1}|c_{0}^{\dagger}\rangle\rangle=t_{F}C_{1}\langle\langle c_{0}|c_{0}^{\dagger}\rangle\rangle+
  t_{F}C_{2}\langle\langle c_{0}^{\dagger}|c_{0}^{\dagger}\rangle\rangle\\
   E_{1+}\langle\langle c_{1}^{\dagger}|c_{0}\rangle\rangle=-t_{F}C_{1}\langle\langle c_{0}^{\dagger}|c_{0}\rangle\rangle-
   t_{F}C_{2}\langle\langle c_{0}|c_{0}\rangle\rangle \\
  E_{1-}\langle\langle c_{1}|c_{0}\rangle\rangle =t_{F}C_{1}\langle\langle c_{0}|c_{0}\rangle\rangle+t_{F}C_{2}\langle\langle c_{0}^{\dagger}|c_{0}\rangle\rangle \\
   E_{1+}\langle\langle c_{1}^{\dagger}|c_{0}^{\dagger}\rangle\rangle=-t_{F}C_{1}\langle\langle c_{0}^{\dagger}|c_{0}^{\dagger}\rangle\rangle-t_{F}C_{2}\langle\langle c_{0}|c_{0}^{\dagger}\rangle\rangle,
\end{eqnarray}
where: $E_{i\mp}=\omega\mp\beta_{i}\epsilon_{i}$ and $\beta_{i}=u_{i}^{2}-v_{i}^{2}$ ($i=0, 1$).

\section{Toy Model $II$: calculation details}\label{toy_model2}
The Majorana Green's function, written in terms of fermionic operators $\langle\langle\lambda_{1}|\lambda_{1}\rangle\rangle=\frac{1}{2}\langle\langle f+f^{\dagger}|f+f^{\dagger}\rangle\rangle$ is calculated from the set of EOMs for Green's functions:
\begin{eqnarray}
  (\omega-\epsilon_{m})\langle\langle f|f^{\dagger}\rangle\rangle =
   1-\tilde{t}_{m}\langle\langle c_{1} |f^{\dagger}\rangle\rangle+\tilde{t}_{m}\langle\langle c_{1}^{\dagger}|f^{\dagger}\rangle\rangle \\
   (\omega-\epsilon_{m})\langle\langle f|f\rangle\rangle =-\tilde{t}_{m}\langle\langle c_{1} |f\rangle\rangle+\tilde{t}_{m}\langle\langle c_{1}^{\dagger}|f \rangle\rangle  \\
   (\omega+\epsilon_{m})\langle\langle f^{\dagger}|f\rangle\rangle=1+\tilde{t}_{m}\langle\langle c_{1}^{\dagger} |f\rangle\rangle-\tilde{t}_{m}\langle\langle c_{1}|f \rangle\rangle  \\
   (\omega+\epsilon_{m})\langle\langle f^{\dagger}|f^{\dagger}\rangle\rangle=
   \tilde{t}_{m}\langle\langle c_{1}^{\dagger} |f^{\dagger}\rangle\rangle-\tilde{t}_{m}\langle\langle c_{1}|f^{\dagger} \rangle\rangle,
  \end{eqnarray}
  and
  \begin{eqnarray}
    E_{1-}\langle\langle c_{1}|f^{\dagger}\rangle\rangle =-\tilde{t}_{m}\langle\langle f|f^{\dagger}\rangle\rangle-\tilde{t}_{m}\langle\langle f^{\dagger}|f^{\dagger}\rangle\rangle\\
    E_{1-}\langle\langle c_{1}|f \rangle\rangle =-\tilde{t}_{m}\langle\langle f|f \rangle\rangle-\tilde{t}_{m}\langle\langle f^{\dagger}|f \rangle\rangle  \\
     E_{1+}\langle\langle c_{1}^{\dagger}|f^{\dagger}\rangle\rangle =\tilde{t}_{m}\langle\langle f^{\dagger}|f^{\dagger}\rangle\rangle+\tilde{t}_{m}\langle\langle f|f^{\dagger}\rangle\rangle  \\
     E_{1+}\langle\langle c_{1}^{\dagger}|f\rangle\rangle=\tilde{t}_{m}\langle\langle f^{\dagger}|f\rangle\rangle+\tilde{t}_{m}\langle\langle f|f\rangle\rangle,
  \end{eqnarray}
  with $E_{1\mp}$ previously defined.
This set of equations is solved exactly yielding:
 \begin{equation}\label{MajoranaGF}
\langle\langle\lambda_{1}|\lambda_{1}\rangle\rangle=\frac{\omega}{\omega^2-\epsilon_{m}^2-\frac{2\tilde{t}_{m}^2\omega^2}{\omega^2-\beta_{1}^{2}\epsilon_{1}^2}}
\end{equation}

It is instructive to consider some simple limits of Eq.~(\ref{MajoranaGF}). For negligible hybridization between Majoranas, $\epsilon_{m}=0$, and an isolated wire, $t_{m}=0$, we obtain $\langle\langle\lambda_{1}|\lambda_{1}\rangle\rangle_{\omega}=1/\omega$, which describes the Majorana resonance located at Fermi energy. For finite hybridization between Majoranas and the wire decoupled from $\epsilon_{1}$ site we obtain the Green's function of the wire:
\begin{equation}
\langle\langle\lambda_{1}|\lambda_{1}\rangle\rangle=\frac{\omega}{\omega^2-\epsilon_{m}^2}=\frac{1/2}{\omega-\epsilon_{m}}+\frac{1/2}{\omega+\epsilon_{m}},
\end{equation}
which has particle and hole resonances with spectral weights of one-half at $\omega=\pm\epsilon_{m}$ of the fermionic state $f$ composed of hybridized $\lambda_{1}$ and $\lambda_{2}$.



\bibliography{libraryPS}

\begin{thebibliography}{95}
\expandafter\ifx\csname natexlab\endcsname\relax\def\natexlab#1{#1}\fi
\expandafter\ifx\csname bibnamefont\endcsname\relax
  \def\bibnamefont#1{#1}\fi
\expandafter\ifx\csname bibfnamefont\endcsname\relax
  \def\bibfnamefont#1{#1}\fi
\expandafter\ifx\csname citenamefont\endcsname\relax
  \def\citenamefont#1{#1}\fi
\expandafter\ifx\csname url\endcsname\relax
  \def\url#1{\texttt{#1}}\fi
\expandafter\ifx\csname urlprefix\endcsname\relax\def\urlprefix{URL }\fi
\providecommand{\bibinfo}[2]{#2}
\providecommand{\eprint}[2][]{\url{#2}}

\bibitem[{\citenamefont{Majorana}(1937)}]{EttoreMajorana1937}
\bibinfo{author}{\bibfnamefont{E.}~\bibnamefont{Majorana}},
  \bibinfo{journal}{Nouvo Cim.} \textbf{\bibinfo{volume}{14}},
  \bibinfo{pages}{171} (\bibinfo{year}{1937}).

\bibitem[{\citenamefont{Sarma et~al.}(2015)\citenamefont{Sarma, Freedman, and
  Nayak}}]{DasSarma2015}
\bibinfo{author}{\bibfnamefont{S.~D.} \bibnamefont{Sarma}},
  \bibinfo{author}{\bibfnamefont{M.}~\bibnamefont{Freedman}}, \bibnamefont{and}
  \bibinfo{author}{\bibfnamefont{C.}~\bibnamefont{Nayak}},
  \bibinfo{journal}{npj Quantum Inf.} \textbf{\bibinfo{volume}{1}},
  \bibinfo{pages}{15001} (\bibinfo{year}{2015}).

\bibitem[{\citenamefont{Aguado}(2017)}]{Aguado2017a}
\bibinfo{author}{\bibfnamefont{R.}~\bibnamefont{Aguado}}, \bibinfo{journal}{La
  Riv. del Nuovo Cim.} \textbf{\bibinfo{volume}{40}}, \bibinfo{pages}{523}
  (\bibinfo{year}{2017}).

\bibitem[{\citenamefont{Beenakker}(2020)}]{Beenakker2019}
\bibinfo{author}{\bibfnamefont{C.~W.} \bibnamefont{Beenakker}},
  \bibinfo{journal}{SciPost Phys. Lect. Notes} \textbf{\bibinfo{volume}{15}}
  (\bibinfo{year}{2020}), \eprint{1907.06497}.

\bibitem[{\citenamefont{Nayak et~al.}(2008)\citenamefont{Nayak, Simon, Stern,
  Freedman, and {Das Sarma}}}]{Nayak2008}
\bibinfo{author}{\bibfnamefont{C.}~\bibnamefont{Nayak}},
  \bibinfo{author}{\bibfnamefont{S.~H.} \bibnamefont{Simon}},
  \bibinfo{author}{\bibfnamefont{A.}~\bibnamefont{Stern}},
  \bibinfo{author}{\bibfnamefont{M.}~\bibnamefont{Freedman}}, \bibnamefont{and}
  \bibinfo{author}{\bibfnamefont{S.}~\bibnamefont{{Das Sarma}}},
  \bibinfo{journal}{Rev. Mod. Phys.} \textbf{\bibinfo{volume}{80}},
  \bibinfo{pages}{1083} (\bibinfo{year}{2008}), \eprint{0707.1889}.

\bibitem[{\citenamefont{Stanescu}(2017)}]{Stanescu2017book}
\bibinfo{author}{\bibfnamefont{T.~D.} \bibnamefont{Stanescu}},
  \emph{\bibinfo{title}{{Introduction to Topolgical Quantum Matter {\&} Quantum
  Computation}}} (\bibinfo{publisher}{CRC Press Taylor {\&} Francis},
  \bibinfo{address}{Boca Raton, London, New York}, \bibinfo{year}{2017}), ISBN
  \bibinfo{isbn}{9781482245936}.

\bibitem[{\citenamefont{Kitaev}(2001)}]{Kitaev2000b}
\bibinfo{author}{\bibfnamefont{A.}~\bibnamefont{Kitaev}},
  \bibinfo{journal}{Phys.-Usp.} \textbf{\bibinfo{volume}{44}},
  \bibinfo{pages}{131} (\bibinfo{year}{2001}), \eprint{0010440}.

\bibitem[{\citenamefont{Lutchyn et~al.}(2010)\citenamefont{Lutchyn, Sau, and
  {Das Sarma}}}]{Lutchyn2010b}
\bibinfo{author}{\bibfnamefont{R.~M.} \bibnamefont{Lutchyn}},
  \bibinfo{author}{\bibfnamefont{J.~D.} \bibnamefont{Sau}}, \bibnamefont{and}
  \bibinfo{author}{\bibfnamefont{S.}~\bibnamefont{{Das Sarma}}},
  \bibinfo{journal}{Phys. Rev. Lett.} \textbf{\bibinfo{volume}{105}},
  \bibinfo{pages}{077001} (\bibinfo{year}{2010}), \eprint{1002.4033}.

\bibitem[{\citenamefont{Oreg et~al.}(2010)\citenamefont{Oreg, Refael, and {Von
  Oppen}}}]{Oreg2010a}
\bibinfo{author}{\bibfnamefont{Y.}~\bibnamefont{Oreg}},
  \bibinfo{author}{\bibfnamefont{G.}~\bibnamefont{Refael}}, \bibnamefont{and}
  \bibinfo{author}{\bibfnamefont{F.}~\bibnamefont{{Von Oppen}}},
  \bibinfo{journal}{Phys. Rev. Lett.} \textbf{\bibinfo{volume}{105}},
  \bibinfo{pages}{177002} (\bibinfo{year}{2010}), \eprint{1003.1145}.

\bibitem[{\citenamefont{Mourik et~al.}(2012)\citenamefont{Mourik, Zuo, Frolov,
  Plissard, Bakkers, and Kouwenhoven}}]{Mourik2012}
\bibinfo{author}{\bibfnamefont{V.}~\bibnamefont{Mourik}},
  \bibinfo{author}{\bibfnamefont{K.}~\bibnamefont{Zuo}},
  \bibinfo{author}{\bibfnamefont{S.~M.} \bibnamefont{Frolov}},
  \bibinfo{author}{\bibfnamefont{S.~R.} \bibnamefont{Plissard}},
  \bibinfo{author}{\bibfnamefont{E.~P. A.~M.} \bibnamefont{Bakkers}},
  \bibnamefont{and} \bibinfo{author}{\bibfnamefont{L.~P.}
  \bibnamefont{Kouwenhoven}}, \bibinfo{journal}{Science}
  \textbf{\bibinfo{volume}{336}}, \bibinfo{pages}{1003} (\bibinfo{year}{2012}),
  \eprint{1204.2792}.

\bibitem[{\citenamefont{Das et~al.}(2012)\citenamefont{Das, Ronen, Most, Oreg,
  Heiblum, and Shtrikman}}]{Das2012}
\bibinfo{author}{\bibfnamefont{A.}~\bibnamefont{Das}},
  \bibinfo{author}{\bibfnamefont{Y.}~\bibnamefont{Ronen}},
  \bibinfo{author}{\bibfnamefont{Y.}~\bibnamefont{Most}},
  \bibinfo{author}{\bibfnamefont{Y.}~\bibnamefont{Oreg}},
  \bibinfo{author}{\bibfnamefont{M.}~\bibnamefont{Heiblum}}, \bibnamefont{and}
  \bibinfo{author}{\bibfnamefont{H.}~\bibnamefont{Shtrikman}},
  \bibinfo{journal}{Nat. Phys.} \textbf{\bibinfo{volume}{8}},
  \bibinfo{pages}{887} (\bibinfo{year}{2012}), \eprint{1205.7073}.

\bibitem[{\citenamefont{Deng et~al.}(2014)\citenamefont{Deng, Yu, Huang,
  Larsson, Caroff, and Xu}}]{Deng2014a}
\bibinfo{author}{\bibfnamefont{M.~T.} \bibnamefont{Deng}},
  \bibinfo{author}{\bibfnamefont{C.~L.} \bibnamefont{Yu}},
  \bibinfo{author}{\bibfnamefont{G.~Y.} \bibnamefont{Huang}},
  \bibinfo{author}{\bibfnamefont{M.}~\bibnamefont{Larsson}},
  \bibinfo{author}{\bibfnamefont{P.}~\bibnamefont{Caroff}}, \bibnamefont{and}
  \bibinfo{author}{\bibfnamefont{H.~Q.} \bibnamefont{Xu}},
  \bibinfo{journal}{Sci. Rep.} \textbf{\bibinfo{volume}{4}},
  \bibinfo{pages}{7261} (\bibinfo{year}{2014}), \eprint{1406.4435}.

\bibitem[{\citenamefont{Higginbotham et~al.}(2015)\citenamefont{Higginbotham,
  Albrecht, Kirsanskas, Chang, Kuemmeth, Krogstrup, Jespersen, Nygard,
  Flensberg, and Marcus}}]{Higginbotham2015}
\bibinfo{author}{\bibfnamefont{A.~P.} \bibnamefont{Higginbotham}},
  \bibinfo{author}{\bibfnamefont{S.~M.} \bibnamefont{Albrecht}},
  \bibinfo{author}{\bibfnamefont{G.}~\bibnamefont{Kirsanskas}},
  \bibinfo{author}{\bibfnamefont{W.}~\bibnamefont{Chang}},
  \bibinfo{author}{\bibfnamefont{F.}~\bibnamefont{Kuemmeth}},
  \bibinfo{author}{\bibfnamefont{P.}~\bibnamefont{Krogstrup}},
  \bibinfo{author}{\bibfnamefont{T.~S.} \bibnamefont{Jespersen}},
  \bibinfo{author}{\bibfnamefont{J.}~\bibnamefont{Nygard}},
  \bibinfo{author}{\bibfnamefont{K.}~\bibnamefont{Flensberg}},
  \bibnamefont{and} \bibinfo{author}{\bibfnamefont{C.~M.}
  \bibnamefont{Marcus}}, \bibinfo{journal}{Nat. Phys.}
  \textbf{\bibinfo{volume}{11}}, \bibinfo{pages}{1017} (\bibinfo{year}{2015}),
  \eprint{1501.05155}.

\bibitem[{\citenamefont{Chang et~al.}(2015)\citenamefont{Chang, Albrecht,
  Jespersen, Kuemmeth, Krogstrup, Nyg{\aa}rd, and Marcus}}]{Chang2015}
\bibinfo{author}{\bibfnamefont{W.}~\bibnamefont{Chang}},
  \bibinfo{author}{\bibfnamefont{S.~M.} \bibnamefont{Albrecht}},
  \bibinfo{author}{\bibfnamefont{T.~S.} \bibnamefont{Jespersen}},
  \bibinfo{author}{\bibfnamefont{F.}~\bibnamefont{Kuemmeth}},
  \bibinfo{author}{\bibfnamefont{P.}~\bibnamefont{Krogstrup}},
  \bibinfo{author}{\bibfnamefont{J.}~\bibnamefont{Nyg{\aa}rd}},
  \bibnamefont{and} \bibinfo{author}{\bibfnamefont{C.~M.}
  \bibnamefont{Marcus}}, \bibinfo{journal}{Nat. Nanotechnol.}
  \textbf{\bibinfo{volume}{10}}, \bibinfo{pages}{232} (\bibinfo{year}{2015}),
  \eprint{1411.6255}.

\bibitem[{\citenamefont{Deng et~al.}(2016)\citenamefont{Deng, Vaitiekenas,
  Hansen, Danon, Leijnse, Flensberg, Nyg{\aa}rd, Krogstrup, and
  Marcus}}]{Deng2016a}
\bibinfo{author}{\bibfnamefont{M.~T.} \bibnamefont{Deng}},
  \bibinfo{author}{\bibfnamefont{S.}~\bibnamefont{Vaitiekenas}},
  \bibinfo{author}{\bibfnamefont{E.~B.} \bibnamefont{Hansen}},
  \bibinfo{author}{\bibfnamefont{J.}~\bibnamefont{Danon}},
  \bibinfo{author}{\bibfnamefont{M.}~\bibnamefont{Leijnse}},
  \bibinfo{author}{\bibfnamefont{K.}~\bibnamefont{Flensberg}},
  \bibinfo{author}{\bibfnamefont{J.}~\bibnamefont{Nyg{\aa}rd}},
  \bibinfo{author}{\bibfnamefont{P.}~\bibnamefont{Krogstrup}},
  \bibnamefont{and} \bibinfo{author}{\bibfnamefont{C.~M.}
  \bibnamefont{Marcus}}, \bibinfo{journal}{Science}
  \textbf{\bibinfo{volume}{354}}, \bibinfo{pages}{1557} (\bibinfo{year}{2016}),
  \eprint{1612.07989}.

\bibitem[{\citenamefont{Chen et~al.}(2017)\citenamefont{Chen, Yu, Stenger,
  Hocevar, Car, Plissard, Bakkers, Stanescu, and Frolov}}]{Chen2017a}
\bibinfo{author}{\bibfnamefont{J.}~\bibnamefont{Chen}},
  \bibinfo{author}{\bibfnamefont{P.}~\bibnamefont{Yu}},
  \bibinfo{author}{\bibfnamefont{J.}~\bibnamefont{Stenger}},
  \bibinfo{author}{\bibfnamefont{M.}~\bibnamefont{Hocevar}},
  \bibinfo{author}{\bibfnamefont{D.}~\bibnamefont{Car}},
  \bibinfo{author}{\bibfnamefont{S.~R.} \bibnamefont{Plissard}},
  \bibinfo{author}{\bibfnamefont{E.~P.} \bibnamefont{Bakkers}},
  \bibinfo{author}{\bibfnamefont{T.~D.} \bibnamefont{Stanescu}},
  \bibnamefont{and} \bibinfo{author}{\bibfnamefont{S.~M.}
  \bibnamefont{Frolov}}, \bibinfo{journal}{Sci. Adv.}
  \textbf{\bibinfo{volume}{3}}, \bibinfo{pages}{1701476}
  (\bibinfo{year}{2017}).

\bibitem[{\citenamefont{Moor et~al.}(2018)\citenamefont{Moor, Bommer, Xu,
  Winkler, Antipov, Bargerbos, Wang, Loo, Op, Gazibegovic et~al.}}]{Moor2018}
\bibinfo{author}{\bibfnamefont{M.~W. A.~D.} \bibnamefont{Moor}},
  \bibinfo{author}{\bibfnamefont{J.~D.~S.} \bibnamefont{Bommer}},
  \bibinfo{author}{\bibfnamefont{D.}~\bibnamefont{Xu}},
  \bibinfo{author}{\bibfnamefont{G.~W.} \bibnamefont{Winkler}},
  \bibinfo{author}{\bibfnamefont{A.~E.} \bibnamefont{Antipov}},
  \bibinfo{author}{\bibfnamefont{A.}~\bibnamefont{Bargerbos}},
  \bibinfo{author}{\bibfnamefont{G.}~\bibnamefont{Wang}},
  \bibinfo{author}{\bibfnamefont{N.~V.} \bibnamefont{Loo}},
  \bibinfo{author}{\bibfnamefont{R.~L.~M.} \bibnamefont{Op}},
  \bibinfo{author}{\bibfnamefont{S.}~\bibnamefont{Gazibegovic}},
  \bibnamefont{et~al.}, \bibinfo{journal}{New J. Phys.}
  \textbf{\bibinfo{volume}{20}}, \bibinfo{pages}{103049}
  (\bibinfo{year}{2018}).

\bibitem[{\citenamefont{Bommer et~al.}(2019)\citenamefont{Bommer, Zhang,
  G{\"{u}}l, Nijholt, Wimmer, Rybakov, Garaud, Rodic, Babaev, Troyer
  et~al.}}]{Bommer2019}
\bibinfo{author}{\bibfnamefont{J.~D.} \bibnamefont{Bommer}},
  \bibinfo{author}{\bibfnamefont{H.}~\bibnamefont{Zhang}},
  \bibinfo{author}{\bibfnamefont{{\"{O}}.}~\bibnamefont{G{\"{u}}l}},
  \bibinfo{author}{\bibfnamefont{B.}~\bibnamefont{Nijholt}},
  \bibinfo{author}{\bibfnamefont{M.}~\bibnamefont{Wimmer}},
  \bibinfo{author}{\bibfnamefont{F.~N.} \bibnamefont{Rybakov}},
  \bibinfo{author}{\bibfnamefont{J.}~\bibnamefont{Garaud}},
  \bibinfo{author}{\bibfnamefont{D.}~\bibnamefont{Rodic}},
  \bibinfo{author}{\bibfnamefont{E.}~\bibnamefont{Babaev}},
  \bibinfo{author}{\bibfnamefont{M.}~\bibnamefont{Troyer}},
  \bibnamefont{et~al.}, \bibinfo{journal}{Phys. Rev. Lett.}
  \textbf{\bibinfo{volume}{122}}, \bibinfo{pages}{187702}
  (\bibinfo{year}{2019}), \eprint{1807.01940}.

\bibitem[{\citenamefont{Karzig et~al.}(2021)\citenamefont{Karzig, Cole, and
  Pikulin}}]{Karzig2021}
\bibinfo{author}{\bibfnamefont{T.}~\bibnamefont{Karzig}},
  \bibinfo{author}{\bibfnamefont{W.~S.} \bibnamefont{Cole}}, \bibnamefont{and}
  \bibinfo{author}{\bibfnamefont{D.~I.} \bibnamefont{Pikulin}},
  \bibinfo{journal}{Phys. Rev. Lett.} \textbf{\bibinfo{volume}{126}},
  \bibinfo{pages}{057702} (\bibinfo{year}{2021}), \eprint{2004.01264}.

\bibitem[{\citenamefont{Gebhard}(1997)}]{Gebhard1997}
\bibinfo{author}{\bibfnamefont{F.}~\bibnamefont{Gebhard}},
  \emph{\bibinfo{title}{{The Mott Metal-Insulator Transition}}}
  (\bibinfo{publisher}{Springer-Verlag Berlin Heidelberg},
  \bibinfo{address}{Berlin, Heidelberg}, \bibinfo{year}{1997}), ISBN
  \bibinfo{isbn}{978-3-540-61481-4}.

\bibitem[{\citenamefont{Hewson}(1993)}]{Hewson1993}
\bibinfo{author}{\bibfnamefont{A.~C.} \bibnamefont{Hewson}},
  \emph{\bibinfo{title}{{The Kondo Problem to Heavy Fermions}}}
  (\bibinfo{publisher}{Cambridge University Press}, \bibinfo{year}{1993}), ISBN
  \bibinfo{isbn}{9780511470752}.

\bibitem[{\citenamefont{Kastner et~al.}(1998)\citenamefont{Kastner,
  Goldhaber-Gordon, Shtrikman, Mahalu, Abusch-Magder, and
  Meirav}}]{Kastner1998}
\bibinfo{author}{\bibfnamefont{M.~A.} \bibnamefont{Kastner}},
  \bibinfo{author}{\bibfnamefont{D.}~\bibnamefont{Goldhaber-Gordon}},
  \bibinfo{author}{\bibfnamefont{H.}~\bibnamefont{Shtrikman}},
  \bibinfo{author}{\bibfnamefont{D.}~\bibnamefont{Mahalu}},
  \bibinfo{author}{\bibfnamefont{D.}~\bibnamefont{Abusch-Magder}},
  \bibnamefont{and} \bibinfo{author}{\bibfnamefont{U.}~\bibnamefont{Meirav}},
  \bibinfo{journal}{Nature} \textbf{\bibinfo{volume}{391}},
  \bibinfo{pages}{156} (\bibinfo{year}{1998}).

\bibitem[{\citenamefont{Gangadharaiah et~al.}(2011)\citenamefont{Gangadharaiah,
  Braunecker, Simon, and Loss}}]{Gangadharaiah2011}
\bibinfo{author}{\bibfnamefont{S.}~\bibnamefont{Gangadharaiah}},
  \bibinfo{author}{\bibfnamefont{B.}~\bibnamefont{Braunecker}},
  \bibinfo{author}{\bibfnamefont{P.}~\bibnamefont{Simon}}, \bibnamefont{and}
  \bibinfo{author}{\bibfnamefont{D.}~\bibnamefont{Loss}},
  \bibinfo{journal}{Phys. Rev. Lett.} \textbf{\bibinfo{volume}{107}},
  \bibinfo{pages}{036801} (\bibinfo{year}{2011}).

\bibitem[{\citenamefont{Katsura et~al.}(2015)\citenamefont{Katsura, Schuricht,
  and Takahashi}}]{Katsura2015}
\bibinfo{author}{\bibfnamefont{H.}~\bibnamefont{Katsura}},
  \bibinfo{author}{\bibfnamefont{D.}~\bibnamefont{Schuricht}},
  \bibnamefont{and}
  \bibinfo{author}{\bibfnamefont{M.}~\bibnamefont{Takahashi}},
  \bibinfo{journal}{Phys. Rev. B} \textbf{\bibinfo{volume}{92}},
  \bibinfo{pages}{115137} (\bibinfo{year}{2015}), \eprint{1507.04444}.

\bibitem[{\citenamefont{Ng}(2015)}]{Ng2015b}
\bibinfo{author}{\bibfnamefont{H.~T.} \bibnamefont{Ng}}, \bibinfo{journal}{Sci.
  Rep.} \textbf{\bibinfo{volume}{5}}, \bibinfo{pages}{12530}
  (\bibinfo{year}{2015}), \eprint{1409.6102}.

\bibitem[{\citenamefont{Herviou et~al.}(2016)\citenamefont{Herviou, Mora, and
  {Le Hur}}}]{Herviou2016a}
\bibinfo{author}{\bibfnamefont{L.}~\bibnamefont{Herviou}},
  \bibinfo{author}{\bibfnamefont{C.}~\bibnamefont{Mora}}, \bibnamefont{and}
  \bibinfo{author}{\bibfnamefont{K.}~\bibnamefont{{Le Hur}}},
  \bibinfo{journal}{Phys. Rev. B} \textbf{\bibinfo{volume}{93}},
  \bibinfo{pages}{165142} (\bibinfo{year}{2016}), \eprint{1601.02998}.

\bibitem[{\citenamefont{Miao et~al.}(2017)\citenamefont{Miao, Jin, Zhang, and
  Zhou}}]{Miao2017a}
\bibinfo{author}{\bibfnamefont{J.~J.} \bibnamefont{Miao}},
  \bibinfo{author}{\bibfnamefont{H.~K.} \bibnamefont{Jin}},
  \bibinfo{author}{\bibfnamefont{F.~C.} \bibnamefont{Zhang}}, \bibnamefont{and}
  \bibinfo{author}{\bibfnamefont{Y.}~\bibnamefont{Zhou}},
  \bibinfo{journal}{Phys. Rev. Lett.} \textbf{\bibinfo{volume}{118}},
  \bibinfo{pages}{267701} (\bibinfo{year}{2017}).

\bibitem[{\citenamefont{Sekania et~al.}(2017)\citenamefont{Sekania, Plugge,
  Greiter, Thomale, and Schmitteckert}}]{Sekania2017}
\bibinfo{author}{\bibfnamefont{M.}~\bibnamefont{Sekania}},
  \bibinfo{author}{\bibfnamefont{S.}~\bibnamefont{Plugge}},
  \bibinfo{author}{\bibfnamefont{M.}~\bibnamefont{Greiter}},
  \bibinfo{author}{\bibfnamefont{R.}~\bibnamefont{Thomale}}, \bibnamefont{and}
  \bibinfo{author}{\bibfnamefont{P.}~\bibnamefont{Schmitteckert}},
  \bibinfo{journal}{Phys. Rev. B} \textbf{\bibinfo{volume}{96}},
  \bibinfo{pages}{094307} (\bibinfo{year}{2017}), \eprint{1703.03360}.

\bibitem[{\citenamefont{Ezawa}(2017)}]{Ezawa2017}
\bibinfo{author}{\bibfnamefont{M.}~\bibnamefont{Ezawa}},
  \bibinfo{journal}{Phys. Rev. B} \textbf{\bibinfo{volume}{96}},
  \bibinfo{pages}{121105(R)} (\bibinfo{year}{2017}), \eprint{1707.03983}.

\bibitem[{\citenamefont{Li and Han}(2018)}]{Li2018c}
\bibinfo{author}{\bibfnamefont{Z.}~\bibnamefont{Li}} \bibnamefont{and}
  \bibinfo{author}{\bibfnamefont{Q.}~\bibnamefont{Han}},
  \bibinfo{journal}{Chinese Phys. Lett.} \textbf{\bibinfo{volume}{35}},
  \bibinfo{pages}{047101} (\bibinfo{year}{2018}), \eprint{1805.01632}.

\bibitem[{\citenamefont{Miao et~al.}(2018)\citenamefont{Miao, Jin, Zhang, and
  Zhou}}]{Miao2018}
\bibinfo{author}{\bibfnamefont{J.~J.} \bibnamefont{Miao}},
  \bibinfo{author}{\bibfnamefont{H.~K.} \bibnamefont{Jin}},
  \bibinfo{author}{\bibfnamefont{F.~C.} \bibnamefont{Zhang}}, \bibnamefont{and}
  \bibinfo{author}{\bibfnamefont{Y.}~\bibnamefont{Zhou}},
  \bibinfo{journal}{Sci. Rep.} \textbf{\bibinfo{volume}{8}},
  \bibinfo{pages}{488} (\bibinfo{year}{2018}), \eprint{1608.08382}.

\bibitem[{\citenamefont{Sarkar}(2020)}]{Sarkar2020}
\bibinfo{author}{\bibfnamefont{S.}~\bibnamefont{Sarkar}},
  \bibinfo{journal}{Sci. Rep.} \textbf{\bibinfo{volume}{10}},
  \bibinfo{pages}{2299} (\bibinfo{year}{2020}).

\bibitem[{\citenamefont{Zvyagin}(2021)}]{Zvyagin2021}
\bibinfo{author}{\bibfnamefont{A.~A.} \bibnamefont{Zvyagin}},
  \bibinfo{journal}{Phys. Rev. B} \textbf{\bibinfo{volume}{103}},
  \bibinfo{pages}{205136 (2021)} (\bibinfo{year}{2021}).

\bibitem[{\citenamefont{Stoudenmire et~al.}(2011)\citenamefont{Stoudenmire,
  Alicea, Starykh, and Fisher}}]{Stoudenmire2011b}
\bibinfo{author}{\bibfnamefont{E.~M.} \bibnamefont{Stoudenmire}},
  \bibinfo{author}{\bibfnamefont{J.}~\bibnamefont{Alicea}},
  \bibinfo{author}{\bibfnamefont{O.~A.} \bibnamefont{Starykh}},
  \bibnamefont{and} \bibinfo{author}{\bibfnamefont{M.~P.}
  \bibnamefont{Fisher}}, \bibinfo{journal}{Phys. Rev. B}
  \textbf{\bibinfo{volume}{84}}, \bibinfo{pages}{014503}
  (\bibinfo{year}{2011}), \eprint{1104.5493}.

\bibitem[{\citenamefont{Lutchyn and Fisher}(2011)}]{Lutchyn2011}
\bibinfo{author}{\bibfnamefont{R.~M.} \bibnamefont{Lutchyn}} \bibnamefont{and}
  \bibinfo{author}{\bibfnamefont{M.~P.} \bibnamefont{Fisher}},
  \bibinfo{journal}{Phys. Rev. B - Condens. Matter Mater. Phys.}
  \textbf{\bibinfo{volume}{84}}, \bibinfo{pages}{214528}
  (\bibinfo{year}{2011}), \eprint{1104.2358}.

\bibitem[{\citenamefont{Klinovaja et~al.}(2012)\citenamefont{Klinovaja, Stano,
  and Loss}}]{Klinovaja2012}
\bibinfo{author}{\bibfnamefont{J.}~\bibnamefont{Klinovaja}},
  \bibinfo{author}{\bibfnamefont{P.}~\bibnamefont{Stano}}, \bibnamefont{and}
  \bibinfo{author}{\bibfnamefont{D.}~\bibnamefont{Loss}},
  \bibinfo{journal}{Phys. Rev. Lett.} \textbf{\bibinfo{volume}{109}},
  \bibinfo{pages}{236801} (\bibinfo{year}{2012}).

\bibitem[{\citenamefont{Maier et~al.}(2014)\citenamefont{Maier, Meng, and
  Loss}}]{Maier2014}
\bibinfo{author}{\bibfnamefont{F.}~\bibnamefont{Maier}},
  \bibinfo{author}{\bibfnamefont{T.}~\bibnamefont{Meng}}, \bibnamefont{and}
  \bibinfo{author}{\bibfnamefont{D.}~\bibnamefont{Loss}},
  \bibinfo{journal}{Phys. Rev. B} \textbf{\bibinfo{volume}{90}},
  \bibinfo{pages}{155437} (\bibinfo{year}{2014}), \eprint{1408.0631}.

\bibitem[{\citenamefont{Meidan et~al.}(2014)\citenamefont{Meidan, Romito, and
  Brouwer}}]{Meidan2014}
\bibinfo{author}{\bibfnamefont{D.}~\bibnamefont{Meidan}},
  \bibinfo{author}{\bibfnamefont{A.}~\bibnamefont{Romito}}, \bibnamefont{and}
  \bibinfo{author}{\bibfnamefont{P.~W.} \bibnamefont{Brouwer}},
  \bibinfo{journal}{Phys. Rev. Lett.} \textbf{\bibinfo{volume}{113}},
  \bibinfo{pages}{057003} (\bibinfo{year}{2014}), \eprint{1312.6367}.

\bibitem[{\citenamefont{Manolescu et~al.}(2014)\citenamefont{Manolescu,
  Marinescu, and Stanescu}}]{Manolescu2014}
\bibinfo{author}{\bibfnamefont{A.}~\bibnamefont{Manolescu}},
  \bibinfo{author}{\bibfnamefont{D.~C.} \bibnamefont{Marinescu}},
  \bibnamefont{and} \bibinfo{author}{\bibfnamefont{T.~D.}
  \bibnamefont{Stanescu}}, \bibinfo{journal}{J. Phys. Condens. Matter}
  \textbf{\bibinfo{volume}{26}}, \bibinfo{pages}{172203}
  (\bibinfo{year}{2014}), \eprint{1312.3888}.

\bibitem[{\citenamefont{Kells}(2015)}]{Kells2015b}
\bibinfo{author}{\bibfnamefont{G.}~\bibnamefont{Kells}},
  \bibinfo{journal}{Phys. Rev. B} \textbf{\bibinfo{volume}{92}},
  \bibinfo{pages}{155434} (\bibinfo{year}{2015}), \eprint{1507.06539}.

\bibitem[{\citenamefont{Chan et~al.}(2015)\citenamefont{Chan, Chiu, and
  Sun}}]{Chan2015}
\bibinfo{author}{\bibfnamefont{Y.~H.} \bibnamefont{Chan}},
  \bibinfo{author}{\bibfnamefont{C.~K.} \bibnamefont{Chiu}}, \bibnamefont{and}
  \bibinfo{author}{\bibfnamefont{K.}~\bibnamefont{Sun}},
  \bibinfo{journal}{Phys. Rev. B} \textbf{\bibinfo{volume}{92}},
  \bibinfo{pages}{104514} (\bibinfo{year}{2015}), \eprint{1506.07860}.

\bibitem[{\citenamefont{Zhang and Tian}(2015)}]{Zhang2015a}
\bibinfo{author}{\bibfnamefont{D.~P.} \bibnamefont{Zhang}} \bibnamefont{and}
  \bibinfo{author}{\bibfnamefont{G.~S.} \bibnamefont{Tian}},
  \bibinfo{journal}{Chinese Phys. B} \textbf{\bibinfo{volume}{24}},
  \bibinfo{pages}{080401} (\bibinfo{year}{2015}).

\bibitem[{\citenamefont{Schmidt and Pedder}(2016)}]{Schmidt2016}
\bibinfo{author}{\bibfnamefont{T.~L.} \bibnamefont{Schmidt}} \bibnamefont{and}
  \bibinfo{author}{\bibfnamefont{C.~J.} \bibnamefont{Pedder}},
  \bibinfo{journal}{Phys. Rev. B} \textbf{\bibinfo{volume}{94}},
  \bibinfo{pages}{125420} (\bibinfo{year}{2016}), \eprint{1604.07720}.

\bibitem[{\citenamefont{Xu et~al.}(2016)\citenamefont{Xu, Xiong, and
  Wang}}]{Xu2016}
\bibinfo{author}{\bibfnamefont{H.}~\bibnamefont{Xu}},
  \bibinfo{author}{\bibfnamefont{Y.}~\bibnamefont{Xiong}}, \bibnamefont{and}
  \bibinfo{author}{\bibfnamefont{J.}~\bibnamefont{Wang}},
  \bibinfo{journal}{Phys. Lett. Sect. A Gen. At. Solid State Phys.}
  \textbf{\bibinfo{volume}{380}}, \bibinfo{pages}{3534} (\bibinfo{year}{2016}).

\bibitem[{\citenamefont{Thakurathi et~al.}(2019)\citenamefont{Thakurathi,
  Aseev, Loss, and Klinovaja}}]{Thakurathi2020}
\bibinfo{author}{\bibfnamefont{M.}~\bibnamefont{Thakurathi}},
  \bibinfo{author}{\bibfnamefont{P.~P.} \bibnamefont{Aseev}},
  \bibinfo{author}{\bibfnamefont{D.}~\bibnamefont{Loss}}, \bibnamefont{and}
  \bibinfo{author}{\bibfnamefont{J.}~\bibnamefont{Klinovaja}},
  \bibinfo{journal}{Phys. Rev. Res.} \textbf{\bibinfo{volume}{2}},
  \bibinfo{pages}{013292} (\bibinfo{year}{2019}), \eprint{1910.03730}.

\bibitem[{\citenamefont{Rylands}(2020)}]{Rylands2020}
\bibinfo{author}{\bibfnamefont{C.}~\bibnamefont{Rylands}},
  \bibinfo{journal}{Phys. Rev. B} \textbf{\bibinfo{volume}{101}},
  \bibinfo{pages}{085133} (\bibinfo{year}{2020}).

\bibitem[{\citenamefont{Wang et~al.}(2020)\citenamefont{Wang, Furusaki, and
  Starykh}}]{Wang2020}
\bibinfo{author}{\bibfnamefont{R.~B.} \bibnamefont{Wang}},
  \bibinfo{author}{\bibfnamefont{A.}~\bibnamefont{Furusaki}}, \bibnamefont{and}
  \bibinfo{author}{\bibfnamefont{O.~A.} \bibnamefont{Starykh}},
  \bibinfo{journal}{Phys. Rev. B} \textbf{\bibinfo{volume}{102}},
  \bibinfo{pages}{165147} (\bibinfo{year}{2020}), \eprint{2007.08482}.

\bibitem[{\citenamefont{Mahyaeh and Ardonne}(2020)}]{Mahyaeh2020}
\bibinfo{author}{\bibfnamefont{I.}~\bibnamefont{Mahyaeh}} \bibnamefont{and}
  \bibinfo{author}{\bibfnamefont{E.}~\bibnamefont{Ardonne}},
  \bibinfo{journal}{Phys. Rev. B} \textbf{\bibinfo{volume}{101}},
  \bibinfo{pages}{85125} (\bibinfo{year}{2020}), \eprint{1911.03156}.

\bibitem[{\citenamefont{Aksenov et~al.}(2020)\citenamefont{Aksenov, Zlotnikov,
  and Shustin}}]{Aksenov2020}
\bibinfo{author}{\bibfnamefont{S.~V.} \bibnamefont{Aksenov}},
  \bibinfo{author}{\bibfnamefont{A.~O.} \bibnamefont{Zlotnikov}},
  \bibnamefont{and} \bibinfo{author}{\bibfnamefont{M.~S.}
  \bibnamefont{Shustin}}, \bibinfo{journal}{Phys. Rev. B}
  \textbf{\bibinfo{volume}{101}}, \bibinfo{pages}{125431}
  (\bibinfo{year}{2020}).

\bibitem[{\citenamefont{H{\"{u}}tzen et~al.}(2012)\citenamefont{H{\"{u}}tzen,
  Zazunov, Braunecker, Yeyati, and Egger}}]{Hutzen2012a}
\bibinfo{author}{\bibfnamefont{R.}~\bibnamefont{H{\"{u}}tzen}},
  \bibinfo{author}{\bibfnamefont{A.}~\bibnamefont{Zazunov}},
  \bibinfo{author}{\bibfnamefont{B.}~\bibnamefont{Braunecker}},
  \bibinfo{author}{\bibfnamefont{A.~L.} \bibnamefont{Yeyati}},
  \bibnamefont{and} \bibinfo{author}{\bibfnamefont{R.}~\bibnamefont{Egger}},
  \bibinfo{journal}{Phys. Rev. Lett.} \textbf{\bibinfo{volume}{109}},
  \bibinfo{pages}{166403} (\bibinfo{year}{2012}), \eprint{1206.3912}.

\bibitem[{\citenamefont{Vijay and Fu}(2016)}]{Vijay2016}
\bibinfo{author}{\bibfnamefont{S.}~\bibnamefont{Vijay}} \bibnamefont{and}
  \bibinfo{author}{\bibfnamefont{L.}~\bibnamefont{Fu}}, \bibinfo{journal}{Phys.
  Rev. B} \textbf{\bibinfo{volume}{94}}, \bibinfo{pages}{235446}
  (\bibinfo{year}{2016}), \eprint{1609.00950}.

\bibitem[{\citenamefont{L{\"{u}} et~al.}(2016)\citenamefont{L{\"{u}}, Lu, and
  Shen}}]{Lu2016}
\bibinfo{author}{\bibfnamefont{H.~F.} \bibnamefont{L{\"{u}}}},
  \bibinfo{author}{\bibfnamefont{H.~Z.} \bibnamefont{Lu}}, \bibnamefont{and}
  \bibinfo{author}{\bibfnamefont{S.~Q.} \bibnamefont{Shen}},
  \bibinfo{journal}{Phys. Rev. B} \textbf{\bibinfo{volume}{93}},
  \bibinfo{pages}{245418} (\bibinfo{year}{2016}).

\bibitem[{\citenamefont{Lutchyn and Glazman}(2017)}]{Lutchyn2017}
\bibinfo{author}{\bibfnamefont{R.~M.} \bibnamefont{Lutchyn}} \bibnamefont{and}
  \bibinfo{author}{\bibfnamefont{L.~I.} \bibnamefont{Glazman}},
  \bibinfo{journal}{Phys. Rev. Lett.} \textbf{\bibinfo{volume}{119}},
  \bibinfo{pages}{057002} (\bibinfo{year}{2017}), \eprint{1701.00184}.

\bibitem[{\citenamefont{Chiu et~al.}(2017)\citenamefont{Chiu, Sau, and {Das
  Sarma}}}]{Chiu2017}
\bibinfo{author}{\bibfnamefont{C.~K.} \bibnamefont{Chiu}},
  \bibinfo{author}{\bibfnamefont{J.~D.} \bibnamefont{Sau}}, \bibnamefont{and}
  \bibinfo{author}{\bibfnamefont{S.}~\bibnamefont{{Das Sarma}}},
  \bibinfo{journal}{Phys. Rev. B} \textbf{\bibinfo{volume}{96}},
  \bibinfo{pages}{054504} (\bibinfo{year}{2017}), \eprint{1702.04357}.

\bibitem[{\citenamefont{Hell et~al.}(2018)\citenamefont{Hell, Flensberg, and
  Leijnse}}]{Hell2018}
\bibinfo{author}{\bibfnamefont{M.}~\bibnamefont{Hell}},
  \bibinfo{author}{\bibfnamefont{K.}~\bibnamefont{Flensberg}},
  \bibnamefont{and} \bibinfo{author}{\bibfnamefont{M.}~\bibnamefont{Leijnse}},
  \bibinfo{journal}{Phys. Rev. B} \textbf{\bibinfo{volume}{97}},
  \bibinfo{pages}{161401(R)} (\bibinfo{year}{2018}), \eprint{1710.05294}.

\bibitem[{\citenamefont{Glazman et~al.}(2019)\citenamefont{Glazman, Pikulin,
  Lutchyn, Flensberg, and Houzet}}]{Glazman2019}
\bibinfo{author}{\bibfnamefont{L.~I.} \bibnamefont{Glazman}},
  \bibinfo{author}{\bibfnamefont{D.}~\bibnamefont{Pikulin}},
  \bibinfo{author}{\bibfnamefont{R.~M.} \bibnamefont{Lutchyn}},
  \bibinfo{author}{\bibfnamefont{K.}~\bibnamefont{Flensberg}},
  \bibnamefont{and} \bibinfo{author}{\bibfnamefont{M.}~\bibnamefont{Houzet}},
  \bibinfo{journal}{Phys. Rev. Lett.} \textbf{\bibinfo{volume}{122}},
  \bibinfo{pages}{16801} (\bibinfo{year}{2019}).

\bibitem[{\citenamefont{Haim et~al.}(2016)\citenamefont{Haim, W{\"{o}}lms,
  Berg, Oreg, and Flensberg}}]{Haim2016}
\bibinfo{author}{\bibfnamefont{A.}~\bibnamefont{Haim}},
  \bibinfo{author}{\bibfnamefont{K.}~\bibnamefont{W{\"{o}}lms}},
  \bibinfo{author}{\bibfnamefont{E.}~\bibnamefont{Berg}},
  \bibinfo{author}{\bibfnamefont{Y.}~\bibnamefont{Oreg}}, \bibnamefont{and}
  \bibinfo{author}{\bibfnamefont{K.}~\bibnamefont{Flensberg}},
  \bibinfo{journal}{Phys. Rev. B} \textbf{\bibinfo{volume}{94}},
  \bibinfo{pages}{115124} (\bibinfo{year}{2016}), \eprint{1605.09385}.

\bibitem[{\citenamefont{Li et~al.}(2019)\citenamefont{Li, Burrello, and
  Flensberg}}]{Li2019}
\bibinfo{author}{\bibfnamefont{T.}~\bibnamefont{Li}},
  \bibinfo{author}{\bibfnamefont{M.}~\bibnamefont{Burrello}}, \bibnamefont{and}
  \bibinfo{author}{\bibfnamefont{K.}~\bibnamefont{Flensberg}},
  \bibinfo{journal}{Phys. Rev. B} \textbf{\bibinfo{volume}{100}},
  \bibinfo{pages}{045305} (\bibinfo{year}{2019}), \eprint{1809.09564}.

\bibitem[{\citenamefont{Zhu et~al.}(2021)\citenamefont{Zhu, Li, Han, and
  Wang}}]{Zhu2021}
\bibinfo{author}{\bibfnamefont{H.~S.} \bibnamefont{Zhu}},
  \bibinfo{author}{\bibfnamefont{Z.}~\bibnamefont{Li}},
  \bibinfo{author}{\bibfnamefont{Q.}~\bibnamefont{Han}}, \bibnamefont{and}
  \bibinfo{author}{\bibfnamefont{Z.~D.} \bibnamefont{Wang}},
  \bibinfo{journal}{Phys. Rev. B} \textbf{\bibinfo{volume}{103}},
  \bibinfo{pages}{024514} (\bibinfo{year}{2021}), \eprint{2011.07548}.

\bibitem[{\citenamefont{Miroshnichenko
  et~al.}(2010)\citenamefont{Miroshnichenko, Flach, and
  Kivshar}}]{Miroshnichenko2010}
\bibinfo{author}{\bibfnamefont{A.~E.} \bibnamefont{Miroshnichenko}},
  \bibinfo{author}{\bibfnamefont{S.}~\bibnamefont{Flach}}, \bibnamefont{and}
  \bibinfo{author}{\bibfnamefont{Y.~S.} \bibnamefont{Kivshar}},
  \bibinfo{journal}{Rev. Mod. Phys.} \textbf{\bibinfo{volume}{82}},
  \bibinfo{pages}{2257} (\bibinfo{year}{2010}), \eprint{0902.3014}.

\bibitem[{\citenamefont{Majorana}(1931)}]{Majorana1931}
\bibinfo{author}{\bibfnamefont{E.}~\bibnamefont{Majorana}},
  \bibinfo{journal}{Nuovo Cim.} p.~\bibinfo{pages}{22} (\bibinfo{year}{1931}).

\bibitem[{\citenamefont{Fano}(1961)}]{Fano1961}
\bibinfo{author}{\bibfnamefont{U.}~\bibnamefont{Fano}}, \bibinfo{journal}{Phys.
  Rev.} \textbf{\bibinfo{volume}{124}}, \bibinfo{pages}{1866}
  (\bibinfo{year}{1961}).

\bibitem[{\citenamefont{Vittorini-Orgeas and
  Bianconi}(2009)}]{Vittorini-Orgeas2009}
\bibinfo{author}{\bibfnamefont{A.}~\bibnamefont{Vittorini-Orgeas}}
  \bibnamefont{and} \bibinfo{author}{\bibfnamefont{A.}~\bibnamefont{Bianconi}},
  \bibinfo{journal}{J. Supercond. Nov. Magn.} \textbf{\bibinfo{volume}{22}},
  \bibinfo{pages}{215} (\bibinfo{year}{2009}), \eprint{0812.1551}.

\bibitem[{\citenamefont{Gong et~al.}(2014)\citenamefont{Gong, Zhang, Li, Yi,
  and Zheng}}]{Gong2014}
\bibinfo{author}{\bibfnamefont{W.~J.} \bibnamefont{Gong}},
  \bibinfo{author}{\bibfnamefont{S.~F.} \bibnamefont{Zhang}},
  \bibinfo{author}{\bibfnamefont{Z.~C.} \bibnamefont{Li}},
  \bibinfo{author}{\bibfnamefont{G.}~\bibnamefont{Yi}}, \bibnamefont{and}
  \bibinfo{author}{\bibfnamefont{Y.~S.} \bibnamefont{Zheng}},
  \bibinfo{journal}{Phys. Rev. B} \textbf{\bibinfo{volume}{89}},
  \bibinfo{pages}{245413} (\bibinfo{year}{2014}), \eprint{1309.2374v1}.

\bibitem[{\citenamefont{Dessotti et~al.}(2014)\citenamefont{Dessotti, Ricco,
  {De Souza}, Souza, and Seridonio}}]{Dessotti2014}
\bibinfo{author}{\bibfnamefont{F.~A.} \bibnamefont{Dessotti}},
  \bibinfo{author}{\bibfnamefont{L.~S.} \bibnamefont{Ricco}},
  \bibinfo{author}{\bibfnamefont{M.}~\bibnamefont{{De Souza}}},
  \bibinfo{author}{\bibfnamefont{F.~M.} \bibnamefont{Souza}}, \bibnamefont{and}
  \bibinfo{author}{\bibfnamefont{A.~C.} \bibnamefont{Seridonio}},
  \bibinfo{journal}{J. Appl. Phys.} \textbf{\bibinfo{volume}{116}},
  \bibinfo{pages}{173701} (\bibinfo{year}{2014}), \eprint{1408.0454}.

\bibitem[{\citenamefont{Ueda and Yokoyama}(2014)}]{Ueda2014}
\bibinfo{author}{\bibfnamefont{A.}~\bibnamefont{Ueda}} \bibnamefont{and}
  \bibinfo{author}{\bibfnamefont{T.}~\bibnamefont{Yokoyama}},
  \bibinfo{journal}{Phys. Rev. B} \textbf{\bibinfo{volume}{90}},
  \bibinfo{pages}{081405(R)} (\bibinfo{year}{2014}), \eprint{1403.4146}.

\bibitem[{\citenamefont{Gong et~al.}(2016)\citenamefont{Gong, Gao, Shan, and
  Yi}}]{Gong2016b}
\bibinfo{author}{\bibfnamefont{W.~J.} \bibnamefont{Gong}},
  \bibinfo{author}{\bibfnamefont{Z.}~\bibnamefont{Gao}},
  \bibinfo{author}{\bibfnamefont{W.~F.} \bibnamefont{Shan}}, \bibnamefont{and}
  \bibinfo{author}{\bibfnamefont{G.~Y.} \bibnamefont{Yi}},
  \bibinfo{journal}{Sci. Rep.} \textbf{\bibinfo{volume}{6}},
  \bibinfo{pages}{23033} (\bibinfo{year}{2016}), \eprint{1501.02529v1}.

\bibitem[{\citenamefont{Nesterov et~al.}(2016)\citenamefont{Nesterov, Houzet,
  and Meyer}}]{Nesterov2015}
\bibinfo{author}{\bibfnamefont{K.~N.} \bibnamefont{Nesterov}},
  \bibinfo{author}{\bibfnamefont{M.}~\bibnamefont{Houzet}}, \bibnamefont{and}
  \bibinfo{author}{\bibfnamefont{J.~S.} \bibnamefont{Meyer}},
  \bibinfo{journal}{Phys. Rev. B} \textbf{\bibinfo{volume}{93}},
  \bibinfo{pages}{174502} (\bibinfo{year}{2016}), \eprint{1512.03042}.

\bibitem[{\citenamefont{Bara{\'{n}}ski
  et~al.}(2017)\citenamefont{Bara{\'{n}}ski, Kobia{\l}ka, and
  Doma{\'{n}}ski}}]{Baranski2017}
\bibinfo{author}{\bibfnamefont{J.}~\bibnamefont{Bara{\'{n}}ski}},
  \bibinfo{author}{\bibfnamefont{A.}~\bibnamefont{Kobia{\l}ka}},
  \bibnamefont{and}
  \bibinfo{author}{\bibfnamefont{T.}~\bibnamefont{Doma{\'{n}}ski}},
  \bibinfo{journal}{J. Phys. Condens. Matter} \textbf{\bibinfo{volume}{29}},
  \bibinfo{pages}{075603} (\bibinfo{year}{2017}), \eprint{1608.02465}.

\bibitem[{\citenamefont{Schuray et~al.}(2017)\citenamefont{Schuray, Weithofer,
  and Recher}}]{Schuray2017a}
\bibinfo{author}{\bibfnamefont{A.}~\bibnamefont{Schuray}},
  \bibinfo{author}{\bibfnamefont{L.}~\bibnamefont{Weithofer}},
  \bibnamefont{and} \bibinfo{author}{\bibfnamefont{P.}~\bibnamefont{Recher}},
  \bibinfo{journal}{Phys. Rev. B} \textbf{\bibinfo{volume}{96}},
  \bibinfo{pages}{085417} (\bibinfo{year}{2017}), \eprint{1702.03909}.

\bibitem[{\citenamefont{Ramos-Andrade et~al.}(2019)\citenamefont{Ramos-Andrade,
  Zambrano, and Orellana}}]{Ramos-Andrade2019}
\bibinfo{author}{\bibfnamefont{J.~P.} \bibnamefont{Ramos-Andrade}},
  \bibinfo{author}{\bibfnamefont{D.}~\bibnamefont{Zambrano}}, \bibnamefont{and}
  \bibinfo{author}{\bibfnamefont{P.~A.} \bibnamefont{Orellana}},
  \bibinfo{journal}{Ann. Phys.} \textbf{\bibinfo{volume}{531}},
  \bibinfo{pages}{1800498} (\bibinfo{year}{2019}), \eprint{1812.04433}.

\bibitem[{\citenamefont{Calle et~al.}(2020)\citenamefont{Calle, Pacheco,
  Orellana, and Ot{\'{a}}lora}}]{Calle2020}
\bibinfo{author}{\bibfnamefont{A.~M.} \bibnamefont{Calle}},
  \bibinfo{author}{\bibfnamefont{M.}~\bibnamefont{Pacheco}},
  \bibinfo{author}{\bibfnamefont{P.~A.} \bibnamefont{Orellana}},
  \bibnamefont{and} \bibinfo{author}{\bibfnamefont{J.~A.}
  \bibnamefont{Ot{\'{a}}lora}}, \bibinfo{journal}{Ann. Phys.}
  \textbf{\bibinfo{volume}{532}}, \bibinfo{pages}{1900409}
  (\bibinfo{year}{2020}).

\bibitem[{\citenamefont{Gong et~al.}(2021)\citenamefont{Gong, Dai, Zhang,
  Jiang, and Gong}}]{Gong2021}
\bibinfo{author}{\bibfnamefont{T.}~\bibnamefont{Gong}},
  \bibinfo{author}{\bibfnamefont{X.-F.} \bibnamefont{Dai}},
  \bibinfo{author}{\bibfnamefont{L.-L.} \bibnamefont{Zhang}},
  \bibinfo{author}{\bibfnamefont{C.}~\bibnamefont{Jiang}}, \bibnamefont{and}
  \bibinfo{author}{\bibfnamefont{W.}~\bibnamefont{Gong}}, \bibinfo{journal}{J.
  Phys. Condens. Matter} \textbf{\bibinfo{volume}{33}}, \bibinfo{pages}{215303}
  (\bibinfo{year}{2021}).

\bibitem[{\citenamefont{Huang et~al.}(2014)\citenamefont{Huang, Leijnse,
  Flensberg, and Xu}}]{Huang2014a}
\bibinfo{author}{\bibfnamefont{G.~Y.} \bibnamefont{Huang}},
  \bibinfo{author}{\bibfnamefont{M.}~\bibnamefont{Leijnse}},
  \bibinfo{author}{\bibfnamefont{K.}~\bibnamefont{Flensberg}},
  \bibnamefont{and} \bibinfo{author}{\bibfnamefont{H.~Q.} \bibnamefont{Xu}},
  \bibinfo{journal}{Phys. Rev. B} \textbf{\bibinfo{volume}{90}},
  \bibinfo{pages}{214507} (\bibinfo{year}{2014}).

\bibitem[{\citenamefont{Rainis et~al.}(2013)\citenamefont{Rainis, Trifunovic,
  Klinovaja, and Loss}}]{Rainis2013}
\bibinfo{author}{\bibfnamefont{D.}~\bibnamefont{Rainis}},
  \bibinfo{author}{\bibfnamefont{L.}~\bibnamefont{Trifunovic}},
  \bibinfo{author}{\bibfnamefont{J.}~\bibnamefont{Klinovaja}},
  \bibnamefont{and} \bibinfo{author}{\bibfnamefont{D.}~\bibnamefont{Loss}},
  \bibinfo{journal}{Phys. Rev. B - Condens. Matter Mater. Phys.}
  \textbf{\bibinfo{volume}{87}}, \bibinfo{pages}{024515}
  (\bibinfo{year}{2013}), ISSN \bibinfo{issn}{10980121}, \eprint{1207.5907}.

\bibitem[{\citenamefont{Jeon et~al.}(2017)\citenamefont{Jeon, Xie, Li, Wang,
  Bernevig, and Yazdani}}]{Jeon2017a}
\bibinfo{author}{\bibfnamefont{S.}~\bibnamefont{Jeon}},
  \bibinfo{author}{\bibfnamefont{Y.}~\bibnamefont{Xie}},
  \bibinfo{author}{\bibfnamefont{J.}~\bibnamefont{Li}},
  \bibinfo{author}{\bibfnamefont{Z.}~\bibnamefont{Wang}},
  \bibinfo{author}{\bibfnamefont{B.~A.} \bibnamefont{Bernevig}},
  \bibnamefont{and} \bibinfo{author}{\bibfnamefont{A.}~\bibnamefont{Yazdani}},
  \bibinfo{journal}{Science} \textbf{\bibinfo{volume}{358}},
  \bibinfo{pages}{772} (\bibinfo{year}{2017}).

\bibitem[{\citenamefont{Prada et~al.}(2020)\citenamefont{Prada, San-Jose,
  de~Moor, Geresdi, Lee, Klinovaja, Loss, Nyg{\aa}rd, Aguado, and
  Kouwenhoven}}]{Prada2020}
\bibinfo{author}{\bibfnamefont{E.}~\bibnamefont{Prada}},
  \bibinfo{author}{\bibfnamefont{P.}~\bibnamefont{San-Jose}},
  \bibinfo{author}{\bibfnamefont{M.~W.} \bibnamefont{de~Moor}},
  \bibinfo{author}{\bibfnamefont{A.}~\bibnamefont{Geresdi}},
  \bibinfo{author}{\bibfnamefont{E.~J.} \bibnamefont{Lee}},
  \bibinfo{author}{\bibfnamefont{J.}~\bibnamefont{Klinovaja}},
  \bibinfo{author}{\bibfnamefont{D.}~\bibnamefont{Loss}},
  \bibinfo{author}{\bibfnamefont{J.}~\bibnamefont{Nyg{\aa}rd}},
  \bibinfo{author}{\bibfnamefont{R.}~\bibnamefont{Aguado}}, \bibnamefont{and}
  \bibinfo{author}{\bibfnamefont{L.~P.} \bibnamefont{Kouwenhoven}},
  \bibinfo{journal}{Nat. Rev. Phys.} \textbf{\bibinfo{volume}{2}},
  \bibinfo{pages}{575} (\bibinfo{year}{2020}), \eprint{1911.04512}.

\bibitem[{\citenamefont{Alicea}(2010)}]{Alicea2010}
\bibinfo{author}{\bibfnamefont{J.}~\bibnamefont{Alicea}},
  \bibinfo{journal}{Phys. Rev. B} \textbf{\bibinfo{volume}{81}},
  \bibinfo{pages}{125318} (\bibinfo{year}{2010}), \eprint{0912.2115}.

\bibitem[{\citenamefont{Sau et~al.}(2010)\citenamefont{Sau, Tewari, Lutchyn,
  Stanescu, and {Das Sarma}}}]{Sau2010a}
\bibinfo{author}{\bibfnamefont{J.~D.} \bibnamefont{Sau}},
  \bibinfo{author}{\bibfnamefont{S.}~\bibnamefont{Tewari}},
  \bibinfo{author}{\bibfnamefont{R.~M.} \bibnamefont{Lutchyn}},
  \bibinfo{author}{\bibfnamefont{T.~D.} \bibnamefont{Stanescu}},
  \bibnamefont{and} \bibinfo{author}{\bibfnamefont{S.}~\bibnamefont{{Das
  Sarma}}}, \bibinfo{journal}{Phys. Rev. B} \textbf{\bibinfo{volume}{82}},
  \bibinfo{pages}{214509} (\bibinfo{year}{2010}), \eprint{1006.2829}.

\bibitem[{\citenamefont{Alicea}(2012)}]{Alicea2012b}
\bibinfo{author}{\bibfnamefont{J.}~\bibnamefont{Alicea}},
  \bibinfo{journal}{Reports Prog. Phys.} \textbf{\bibinfo{volume}{75}},
  \bibinfo{pages}{076501} (\bibinfo{year}{2012}), \eprint{1202.1293}.

\bibitem[{\citenamefont{Kittel}(1987)}]{KittelBook}
\bibinfo{author}{\bibfnamefont{C.}~\bibnamefont{Kittel}},
  \emph{\bibinfo{title}{{Quantum Theory of Solids}}}
  (\bibinfo{publisher}{Willey}, \bibinfo{address}{New York, Chichester,
  Brisbane, Toronto, Singapore}, \bibinfo{year}{1987}), ISBN
  \bibinfo{isbn}{978-0-471-62412-7}.

\bibitem[{\citenamefont{Stefa{\'{n}}ski
  et~al.}(2004)\citenamefont{Stefa{\'{n}}ski, Tagliacozzo, and
  Bu{\l}ka}}]{Stefanski2004d}
\bibinfo{author}{\bibfnamefont{P.}~\bibnamefont{Stefa{\'{n}}ski}},
  \bibinfo{author}{\bibfnamefont{A.}~\bibnamefont{Tagliacozzo}},
  \bibnamefont{and} \bibinfo{author}{\bibfnamefont{B.~R.}
  \bibnamefont{Bu{\l}ka}}, \bibinfo{journal}{Phys. Rev. Lett.}
  \textbf{\bibinfo{volume}{93}}, \bibinfo{pages}{186805}
  (\bibinfo{year}{2004}).

\bibitem[{\citenamefont{Stefa{\'{n}}ski}(2003)}]{Stefanski2003b}
\bibinfo{author}{\bibfnamefont{P.}~\bibnamefont{Stefa{\'{n}}ski}},
  \bibinfo{journal}{Solid State Commun.} \textbf{\bibinfo{volume}{128}},
  \bibinfo{pages}{29} (\bibinfo{year}{2003}).

\bibitem[{\citenamefont{Rainis and Loss}(2012)}]{Rainis2012a}
\bibinfo{author}{\bibfnamefont{D.}~\bibnamefont{Rainis}} \bibnamefont{and}
  \bibinfo{author}{\bibfnamefont{D.}~\bibnamefont{Loss}},
  \bibinfo{journal}{Phys. Rev. B - Condens. Matter Mater. Phys.}
  \textbf{\bibinfo{volume}{85}}, \bibinfo{pages}{174533}
  (\bibinfo{year}{2012}), ISSN \bibinfo{issn}{10980121}, \eprint{1204.3326}.

\bibitem[{\citenamefont{Frolov et~al.}(2020)\citenamefont{Frolov, Manfra, and
  Sau}}]{Frolov2020}
\bibinfo{author}{\bibfnamefont{S.~M.} \bibnamefont{Frolov}},
  \bibinfo{author}{\bibfnamefont{M.~J.} \bibnamefont{Manfra}},
  \bibnamefont{and} \bibinfo{author}{\bibfnamefont{J.~D.} \bibnamefont{Sau}},
  \bibinfo{journal}{Nat. Phys.} \textbf{\bibinfo{volume}{16}},
  \bibinfo{pages}{718} (\bibinfo{year}{2020}).

\bibitem[{\citenamefont{Yu et~al.}(2021)\citenamefont{Yu, Chen, Gomanko,
  Badawy, Bakkers, Zuo, Mourik, and Frolov}}]{Yu2021}
\bibinfo{author}{\bibfnamefont{P.}~\bibnamefont{Yu}},
  \bibinfo{author}{\bibfnamefont{J.}~\bibnamefont{Chen}},
  \bibinfo{author}{\bibfnamefont{M.}~\bibnamefont{Gomanko}},
  \bibinfo{author}{\bibfnamefont{G.}~\bibnamefont{Badawy}},
  \bibinfo{author}{\bibfnamefont{E.~P.} \bibnamefont{Bakkers}},
  \bibinfo{author}{\bibfnamefont{K.}~\bibnamefont{Zuo}},
  \bibinfo{author}{\bibfnamefont{V.}~\bibnamefont{Mourik}}, \bibnamefont{and}
  \bibinfo{author}{\bibfnamefont{S.~M.} \bibnamefont{Frolov}},
  \bibinfo{journal}{Nat. Phys.} \textbf{\bibinfo{volume}{17}},
  \bibinfo{pages}{482} (\bibinfo{year}{2021}).

\bibitem[{\citenamefont{Ruby et~al.}(2015)\citenamefont{Ruby, Pientka, Peng,
  {Von Oppen}, Heinrich, and Franke}}]{Ruby2015}
\bibinfo{author}{\bibfnamefont{M.}~\bibnamefont{Ruby}},
  \bibinfo{author}{\bibfnamefont{F.}~\bibnamefont{Pientka}},
  \bibinfo{author}{\bibfnamefont{Y.}~\bibnamefont{Peng}},
  \bibinfo{author}{\bibfnamefont{F.}~\bibnamefont{{Von Oppen}}},
  \bibinfo{author}{\bibfnamefont{B.~W.} \bibnamefont{Heinrich}},
  \bibnamefont{and} \bibinfo{author}{\bibfnamefont{K.~J.}
  \bibnamefont{Franke}}, \bibinfo{journal}{Phys. Rev. Lett.}
  \textbf{\bibinfo{volume}{115}}, \bibinfo{pages}{087001}
  (\bibinfo{year}{2015}), \eprint{1502.05048}.

\bibitem[{\citenamefont{Huang et~al.}(2020)\citenamefont{Huang, Padurariu,
  Senkpiel, Drost, Yeyati, Cuevas, Kubala, Ankerhold, Kern, and
  Ast}}]{Huang2020}
\bibinfo{author}{\bibfnamefont{H.}~\bibnamefont{Huang}},
  \bibinfo{author}{\bibfnamefont{C.}~\bibnamefont{Padurariu}},
  \bibinfo{author}{\bibfnamefont{J.}~\bibnamefont{Senkpiel}},
  \bibinfo{author}{\bibfnamefont{R.}~\bibnamefont{Drost}},
  \bibinfo{author}{\bibfnamefont{A.~L.} \bibnamefont{Yeyati}},
  \bibinfo{author}{\bibfnamefont{J.~C.} \bibnamefont{Cuevas}},
  \bibinfo{author}{\bibfnamefont{B.}~\bibnamefont{Kubala}},
  \bibinfo{author}{\bibfnamefont{J.}~\bibnamefont{Ankerhold}},
  \bibinfo{author}{\bibfnamefont{K.}~\bibnamefont{Kern}}, \bibnamefont{and}
  \bibinfo{author}{\bibfnamefont{C.~R.} \bibnamefont{Ast}},
  \bibinfo{journal}{Nat. Phys.} \textbf{\bibinfo{volume}{16}},
  \bibinfo{pages}{1227} (\bibinfo{year}{2020}), \eprint{1912.08901}.

\bibitem[{\citenamefont{Rubio-Verd{\'{u}}
  et~al.}(2021)\citenamefont{Rubio-Verd{\'{u}}, Zald{\'{i}}var, {\v{Z}}itko,
  and Pascual}}]{Rubio-Verdu2021}
\bibinfo{author}{\bibfnamefont{C.}~\bibnamefont{Rubio-Verd{\'{u}}}},
  \bibinfo{author}{\bibfnamefont{J.}~\bibnamefont{Zald{\'{i}}var}},
  \bibinfo{author}{\bibfnamefont{R.}~\bibnamefont{{\v{Z}}itko}},
  \bibnamefont{and} \bibinfo{author}{\bibfnamefont{J.~I.}
  \bibnamefont{Pascual}}, \bibinfo{journal}{Phys. Rev. Lett.}
  \textbf{\bibinfo{volume}{126}}, \bibinfo{pages}{017001}
  (\bibinfo{year}{2021}).

\bibitem[{\citenamefont{Lo et~al.}(2014)\citenamefont{Lo, Lin, Wang, Lin, and
  Liang}}]{Lo2014}
\bibinfo{author}{\bibfnamefont{S.~T.} \bibnamefont{Lo}},
  \bibinfo{author}{\bibfnamefont{S.~W.} \bibnamefont{Lin}},
  \bibinfo{author}{\bibfnamefont{Y.~T.} \bibnamefont{Wang}},
  \bibinfo{author}{\bibfnamefont{S.~D.} \bibnamefont{Lin}}, \bibnamefont{and}
  \bibinfo{author}{\bibfnamefont{C.~T.} \bibnamefont{Liang}},
  \bibinfo{journal}{Sci. Rep.} \textbf{\bibinfo{volume}{4}},
  \bibinfo{pages}{5438} (\bibinfo{year}{2014}).

\bibitem[{\citenamefont{Desjardins et~al.}(2019)\citenamefont{Desjardins,
  Contamin, Delbecq, Dartiailh, Bruhat, Cubaynes, Viennot, Mallet, Rohart,
  Thiaville et~al.}}]{Desjardins2019}
\bibinfo{author}{\bibfnamefont{M.~M.} \bibnamefont{Desjardins}},
  \bibinfo{author}{\bibfnamefont{L.~C.} \bibnamefont{Contamin}},
  \bibinfo{author}{\bibfnamefont{M.~R.} \bibnamefont{Delbecq}},
  \bibinfo{author}{\bibfnamefont{M.~C.} \bibnamefont{Dartiailh}},
  \bibinfo{author}{\bibfnamefont{L.~E.} \bibnamefont{Bruhat}},
  \bibinfo{author}{\bibfnamefont{T.}~\bibnamefont{Cubaynes}},
  \bibinfo{author}{\bibfnamefont{J.~J.} \bibnamefont{Viennot}},
  \bibinfo{author}{\bibfnamefont{F.}~\bibnamefont{Mallet}},
  \bibinfo{author}{\bibfnamefont{S.}~\bibnamefont{Rohart}},
  \bibinfo{author}{\bibfnamefont{A.}~\bibnamefont{Thiaville}},
  \bibnamefont{et~al.}, \bibinfo{journal}{Nat. Mater.}
  \textbf{\bibinfo{volume}{18}}, \bibinfo{pages}{1060} (\bibinfo{year}{2019}),
  \eprint{1902.07479}.

\bibitem[{\citenamefont{Lee and Fisher}(1981)}]{Lee1981}
\bibinfo{author}{\bibfnamefont{P.~A.} \bibnamefont{Lee}} \bibnamefont{and}
  \bibinfo{author}{\bibfnamefont{D.~S.} \bibnamefont{Fisher}},
  \bibinfo{journal}{Phys. Rev. Lett.} \textbf{\bibinfo{volume}{47}},
  \bibinfo{pages}{882} (\bibinfo{year}{1981}).

\bibitem[{\citenamefont{MacKinnon}(1985)}]{MacKinnon1985}
\bibinfo{author}{\bibfnamefont{A.}~\bibnamefont{MacKinnon}},
  \bibinfo{journal}{Z. Phys. B-Condensed Matter} \textbf{\bibinfo{volume}{59}},
  \bibinfo{pages}{385} (\bibinfo{year}{1985}).

\bibitem[{\citenamefont{Asano}(2001)}]{Asano2001}
\bibinfo{author}{\bibfnamefont{Y.}~\bibnamefont{Asano}},
  \bibinfo{journal}{Phys. Rev. B} \textbf{\bibinfo{volume}{63}},
  \bibinfo{pages}{052512} (\bibinfo{year}{2001}).

\bibitem[{\citenamefont{Potter and Lee}(2011)}]{Potter2011}
\bibinfo{author}{\bibfnamefont{A.~C.} \bibnamefont{Potter}} \bibnamefont{and}
  \bibinfo{author}{\bibfnamefont{P.~A.} \bibnamefont{Lee}},
  \bibinfo{journal}{Phys. Rev. B} \textbf{\bibinfo{volume}{83}},
  \bibinfo{pages}{094525} (\bibinfo{year}{2011}).

\end{thebibliography}
\end{document}